\documentstyle[11pt,aaspp4]{article}

\lefthead{Gadotti \& dos Anjos}
\righthead{Homogenization of the Stellar Population}

\begin{document}

   \title{Homogenization of the Stellar Population along Late--Type \\ Spiral Galaxies\footnote{Based 
          partly on observations made at the Pico dos Dias Observatory (PDO/LNA -- CNPq), 
          Brazil}}

   \author{D. A. Gadotti}
   \affil{Departamento de Astronomia, Instituto Astron\^omico e Geof\'\i sico, Universidade de S\~ao Paulo}
   \affil{Avenida Miguel Stefano, 4200, S\~ao Paulo -- SP, Brasil, CEP 04301-904}
   \affil{dimitri@iagusp.usp.br}

   \and

   \author{S. dos Anjos}
   \affil{Departamento de Astronomia, Instituto Astron\^omico e Geof\'\i sico, Universidade de S\~ao Paulo}
   \affil{Avenida Miguel Stefano, 4200, S\~ao Paulo -- SP, Brasil, CEP 04301-904}
   \affil{sandra@iagusp.usp.br}
   
   \begin{abstract}

   We present a study of the broadband UBV color profiles for 257 Sbc barred and non--barred 
   galaxies, using photoelectric aperture photometry data from the literature. Using robust 
   statistical methods, we have estimated the color gradients of the galaxies, as well as the total 
   and bulge mean colors. A comparative photometric study using CCD images was done. In our sample,
   the color gradients are negative (reddish inward) in approximately 59\% of the objects, are
   almost null in 27\%, and are positive in 14\%, considering only the face--on galaxies, which 
   represents approximately 51\% of the sample. The results do not change, essentially, when we
   include the edge--on galaxies.
     
   As a consequence of this study we have also found that barred galaxies 
   are over--represented among the objects having null or positive gradients, 
   indicating that bars act as a mechanism of homogenization of the stellar population. This 
   effect is more evident in the (U$-$B) color index, although it can also be detected in the
   (B\,$-$V) color.
   
   A correlation between the total and bulge colors was found, which is a consequence of an underlying
   correlation between the colors of bulges and disks found by other authors. Moreover, the mean 
   total color is the same irrespective of the gradient regime, while bulges are bluer in galaxies
   with null or positive gradients, which indicates an increase of the star formation rate
   in the central regions of these objects.
      
   We have also made a quantitative evaluation of the amount of extinction in the center of these galaxies.
   This was done using WFPC2 and NICMOS HST archival data, as well as CCD B, V and I images.
   We show that although the extinction in the V--band can reach values of up to 2
   magnitudes in the central region, it is unlikely that dust plays a fundamental role in global color
   gradients.

   We found no correlation between color and O/H abundance gradients. This result could suggest
   that the color gradients are more sensitive to the age rather than to the metallicity of
   the stellar population. However, the absence of this correlation may be caused by dust extinction.
   We discuss this result considering a picture in which bars are a
   relatively fast recurrent phenomenon.

   These results are not compatible with a pure classical monolithic scenario for bulge and
   disk formation. On the contrary, they favor a scenario where both these components
   are evolving in a correlated process, in which stellar bars play a crucial role.
      
   \end{abstract}
   
   
\section{Introduction}

   Several recent works have been contributed to our understanding of the
   dynamical evolutionary processes related to stellar bars in galaxies (see \cite{fri99} for a review).
   Much of these works point to the possibility that these processes may be related
   to the formation and/or building of galactic bulges, as opposed to a pure monolithic
   scenario (\cite{egg62}) and the hierarchical scenario (e.g., \cite{kau93,kau94,bau96,bou98}).
   
   We know that bars are very easy to form in 
   stellar disks due to non--circular orbits of the stars in the disk, or due
   to instabilities generated by the presence of a companion. In the RC3 (\cite{dev91}), for instance, 
   30\% of the spiral galaxies are strongly barred.
   Recent theoretical studies based on N--body simulations (e.g., \cite{fri95} 
   and references therein) show that, once formed, the stellar bar induces a series of 
   dynamical processes in the host galaxy. Basically, these studies show two routes for
   the formation and/or building of galactic bulges. In the first one, stellar bars
   could collect gas from the outer disk, generating bursts of star formation and a chemical enrichment
   in the central regions. Another possibility is that the stars themselves might be transported
   from the disk to the bulge, through, for instance, the hose mechanism (\cite{too66}), 
   orbital resonances (\cite{com81}) and the onset of irregular stellar orbits
   (e.g., \cite{ber98}). Moreover, \cite{nor96}, among other theoretical works, 
   showed that the central concentration of mass, induced by the bar, could destroy 
   its orbital structure and eventually the bar itself. These authors suggest that the formation of the 
   bar, its dissolution and consequent formation and/or building of the bulge, may be a 
   fast recurrent process (i.e., $\sim 10^{8}$ years). \cite{fri93} suggest that the continuous building of 
   the bulge in a galaxy could actually change its overall morphology. One Sc galaxy,
   for instance, might become in a first step a SBb, and then a Sb,
   giving an evolutionary meaning to the late--type spiral scheme along the Hubble sequence.
      
   From the observational point of view, comparative studies related to the general
   properties of barred and non--barred galaxies seem to give some support to the
   formation and/or building of galactic bulges through this secular evolutionary scenario.
   \cite{kor82} found that triaxial bulges, which are normally associated with bars, 
   rotate faster than bulges of non--barred galaxies. \cite{kor83} 
   showed that bulges of barred galaxies have a central velocity dispersion smaller 
   than the one presented by bulges in non--barred galaxies. \cite{mar94} 
   and \cite{zar94} show that barred galaxies have less pronounced O/H 
   gradients than non--barred galaxies. \cite{sak99} show that barred galaxies 
   present a higher degree of central concentration of CO molecular gas than non--barred galaxies.
   
   Moreover, box--shaped bulges, representing at least 20--30\% of edge--on S0's
   (\cite{des87,sha87}), seem to show this morphology as a consequence of steps in the
   secular dynamical evolutionary processes in bars, as indicated by a series of recent results
   (\cite{kui95,mer99,bur99a,ath99,bur99b,bur99c}).
   
   Studies related to general properties of spirals (e.g., \cite{pel96}) 
   revealed similar broadband colors of the inner disk and bulge; \cite{cou96} 
   and \cite{dej96b} found a correlation between the scale lengths of 
   disks and bulges. These results indicate the existence of an evolutionary 
   connection between these two components (see \cite{wys97} for a review), and have been 
   interpreted as a consequence of the dynamical secular evolutionary scenario.
   
   Another way to obtain clues related to the formation and evolution processes in galaxies is
   through the study of radial color distribution. However, surprisingly, there are few
   statistical works in the literature exploring the broadband colors to study the bulge and 
   the disk components separately. The study of 
   the integrated broadband colors in galaxies have been done to 
   obtain information concerning the stellar population (e.g., \cite{sea73,tin80,fro85,pel89,sil94}), 
   as well as the internal extinction caused 
   by the interstellar dust (e.g., \cite{eva94,pel94}). 
   Exceptions are the works of \cite{dej94} and \cite{dej96a},
   for instance. Such studies certainly bring clues about the bulge formation scenarios.
     
   The main goal of this paper is to compare the color gradients' behaviour in 
   barred and non--barred late--type galaxies and verify if these results are in agreement with the
   predictions from evolutionary processes. A very fast way to verify alternative 
   scenarios for the formation and/or building of bulges, exploring
   the radial color distribution in a statistical point of view, is to use the available data in the
   literature. 
   
   With this objective, we have selected a sample of 257
   Sbc galaxies with broadband colors available in the literature and
   observed through photoelectric aperture photometry (Sect. 2).
   Using robust statistical methods, we estimated, for each galaxy, the color gradient as well 
   as the mean total and bulge characteristic color indices (Sect. 3). Moreover, 
   we have also acquired CCD images for 14 galaxies in the sample (Sect. 4)  
   in order to test the accuracy of our results. In Sect. 5 we present the main results of our analysis
   and, finally, in Sect. 6 we present a general discussion and our main conclusions.

\section{Sample Selection}

   The photoelectric data used in this analysis were extracted from
   the compilation by \cite{lon83} and its supplement (\cite{lon85}).
   Both compilations will hereafter be referred to as LdV83,85, respectively.
   Among other information, the catalogue presents for different galaxies
   the (U$-$B) and (B\,$-$V) aperture
   color indices extracted from the literature. We have selected
   galaxies with Hubble stage index T = 3, 4 or 5, corresponding to morphological
   types Sb, Sbc and Sc, barred and non--barred, and having B$_{T}$ brighter than 14, 
   according to the Third Reference Catalogue (\cite{dev91}; 
   hereafter RC3). This criterion assures that the morphological classification
   is more reliable, since fainter objects are, in general, 
   more difficult to be classified. Nevertheless, it is worth notice that
   several galaxies were distinctly classified in the LdV83,85 and in the
   RC3. Since the rms uncertainty associated with the morphological type
   is of order 2 units (\cite{lah95}), 
   we will consider all galaxies in our sample as belonging to one unique
   mean morphological class (T = 4 $\pm$ 1).
   
   We remark that our choice for galaxies in this specific type range was
   motivated by the fact that these
   are the most luminous objects in the B broadband along the Hubble sequence
   (\cite{van97,rob94}). This is possibly indicating
   that these systems have the highest rate of star formation among spirals.
   Another reason for this choice comes from the observation
   that it is possible that the dynamical evolutionary processes occur mainly in late--type
   spirals rather than in early--type ones (\cite{wys97}).
   
   A first selection of the data was partially done with the electronic version of the 
   catalogue, available at the CDS ({\it Centre de Donn\'ees Astronomiques de Strasbourg}), 
   and resulted in a sample containing 531 objects. In order to have an equally representative 
   set of data, we have removed
   from the sample those objects with less than 5 different color aperture data.
   Thus, we selected only those objects for which a more careful study of the distribution
   of the color indices could be done.
   
   We know that extinction by dust can strongly affect studies of the radial color distribution in
   the U, B and V bands, in particular for late--type galaxies. However, these bands are well 
   suited for the study proposed here, since in these bands we can detect recent star formation, 
   which is a possible consequence of the dynamical secular evolutionary scenario.
   In order to minimize the effects of dust, we have also 
   performed a visual inspection of all
   galaxies, using images of the DSS (Digitized Sky Survey), eliminating those
   peculiar systems, (e.g., NGC 891), presenting clear perturbations, such as strong dustlanes or close
   companions in strong interaction, that could disturb the analysis. After this last step
   we ended with a final sample having 257 galaxies, used in the present analysis.
      
\section{Estimating Gradients and Colors}

   \subsection{Color Gradients}

   As mentioned before, we have used the LdV83,85 data to estimate the (U$-$B) and (B\,$-$V) 
   color gradients of the galaxies in our sample. Since this is a compilation of data acquired by 
   different observers,
   telescopes, instruments and in different atmospheric conditions, it is natural
   that, for any given galaxy, some data will not appear consistent, due to larger 
   internal errors. For instance,
   different authors could assign quite distinct values for the color index
   of the same galaxy at the same aperture. Indeed, this is the case, for example, of NGC 2377
   in the aperture of 2.6 arcminutes, where three different sources gave to the (U$-$B) 
   color index the values 0.11, 0.20 and 0.38! 
   Therefore, trying to fit a straight line to
   the color data, using 
   these discrepant values and the classical least squares
   regression (LS), will result in a quite uncertain estimative for the color gradient.
   
   Since we do not know a priori how to identify the bad data, it is mandatory to use some 
   robust statistical technique more insensitive to the presence of these
   uncertainties. In our analysis, we choose to apply the Least Median of Squares (LMS) method
   (\cite{rou84}). Contrary to the classical LS regression, this
   method minimizes the {\em median} of the
   squared residuals. The results obtained are more resistant to 
   the effects of contamination in the data. More specifically, the estimation of
   the color gradient was done with the program {\sc progress} (\cite{rou87}), 
   available at the StatLib (http://www.lib.stat.cmu.edu/).
   This program performs a robust
   regression analysis by means of the LMS method, yielding more reliable
   estimates of the regression parameters, and allowing to identify outliers in the data.
   {\sc progress} first calculates the regression parameters by LS, then by
   LMS, and finally by a reweighted LS (in which the outliers
   have weight zero). Through this algorithm, the estimated gradient has, in most cases, 
   the same value obtained through the LMS method alone. However, the reweighted LS method works
   better than the LMS method when the number of data points is small (\cite{rou84,rou87}).
   
   The color gradient was estimated following the same definition of \cite{pru98}, i.e., 
      
   \begin{equation}
   {G} = {{\Delta (X-Y)} \over {\Delta \log {A}}},
   \end{equation}
   
   \noindent where $(X-Y)$ represents the integrated color index in magnitudes within
   an aperture $A$ in units of 0.1 arcminute.

   The estimation of each gradient was also accompanied by a graphical visual
   inspection, since, in some cases, the 
   results from the non--reweighted LS were more representative than those by LMS or by the
   reweighted LS. This could
   happen because, when trying to minimize the errors, the LMS method can be fooled by a small
   group of data points that fits very well a straight line. Thus, in these cases, we defined the
   gradient by the parameters obtained by the classical LS regression.
   
   From our sample of 257 galaxies, we obtained 239 (B\,$-$V) and 202 (U$-$B) color gradients.
   The other estimations were rejected either because the number of data points were
   too small and/or the points were too inconsistent to result in a reliable value. Figure 1
   shows four examples of the radial color distribution in galaxies. 
   NGC 1425 and NGC 2613 are examples of objects having the more typical 
   negative color gradient. An example of object with a clear null gradient is NGC 1672.
   The more rare case of objects with a positive
   gradient is represented here by UGC 3973. 
   In this figure, we can also compare
   the fits using the three different methods discussed above. 
   The dashed lines refer to the standard LS
   method, while dotted lines refer to the LMS method and the solid lines refer to the {\sc progress}
   algorithm. Note the importance of using a robust statistical method
   to determine color gradients in such cases as for NGC 2613 and UGC 3973.
  
\placefigure{fig1}
  
   The LdV83,85 data is not corrected for either Galactic reddening or internal reddening.
   In determining the color gradients, the correction for Galactic reddening is not necessary, since
   it only introduces a constant vertical shift of the points, 
   not affecting the gradient evaluation. However, the 
   correction for internal reddening is quite difficult to predict correctly, due to the still unsolved
   problems related to the optical thickness and the inclination of galaxies
   (e.g., \cite{gio95,dej96c}). However, although such a correction
   could be important for any particular object, 
   it will only produce minor changes compared to the uncertainties involved in the 
   measurement and determination of the color gradients. 
   
   On the other hand, in models of the dust distribution in disk dominated galaxies, it was shown
   (\cite{dej96c}) that only a small fraction of the color gradients could be due
   to dust reddening, i.e., dust reddening plays a minor role in color gradients. Furthermore, 
   color gradients induced by dust are small from U to R broadbands, because the absorption
   properties do not change very much among these bands.
   
   In Fig. 2 we show the color gradients plotted against both the Galactic reddening and the 
   inclination of the galaxies. We can see from this figure that the two corrections mentioned above
   do not interfere with the distribution of color gradients
   obtained in our sample. The top panels show that there is
   no correlation between color gradients and Galactic reddening, represented by the color
   excess $E(B-V)$, determined through the recently obtained maps of \cite{sch98}.
   On the other hand, since the internal reddening varies with the inclination of the galaxy
   along the line of sight, which can be represented by the $\log R_{25}$ parameter of the RC3, 
   the bottom panels of Fig. 2 show that there is no clear correlation between color
   gradients and internal reddening. Since no
   correlation was found, we opt to neglect the internal reddening when estimating
   the color gradients. We remark, however, that both effects are still obviously 
   relevant when dealing with the integrated color, as we present in the next subsection.
  
\placefigure{fig2}
   
   \subsection{Total and Bulge Colors}
   
   We have used two different procedures to determine the total and the bulge
   characteristic color indices. In the first one, we adopt the bulge color 
   as the one observed through
   the smallest aperture, and the total color as the one observed through the 
   aperture that reaches the 25 mag arcsec$^{-2}$ B isophotal level,  
   as presented in the RC3. In some cases, when
   the data do not reach the dimensions required, an extrapolation was done using the 
   estimated gradient. On the other hand, an average was done for apertures with
   several data points. No reddening corrections were made in this method. 
   
   We stress that this method is completely unbiased, in the sense that we use the
   original data, and therefore it is useful to
   verify if the total colors and the colors of bulges are correlated. 
   Such correlation should in fact exist, since the colors of bulges and disks are correlated
   (\cite{pel96}). They have used the (U$-$R), (B$-$R), 
   (R$-$K) and (J$-$K) colors in a sample of 30 early--type spirals (earlier than Sbc).
   As shown in Sect. 5.5, we also found a good correlation, consistent with the findings of 
   these authors.
   
   Since galaxies have different angular sizes and were observed through different sets of
   apertures, the method described above is only an approximated procedure, since, 
   in many cases, the measurements were made at different galactocentric distances.
   In order to compare the data of galaxies at the same physical dimension, we have defined a
   characteristic bulge color index as the one measured within 1/5 of the galaxy effective radius.
   Even if there is some disk contamination at this aperture, the major contribution comes
   from the bulge, and therefore we made no attempt to correct for this contamination.
   Using the definition of gradient (Eq. (1)), this bulge color was derived from our fits as
   
   \begin{equation}
   {{(X-Y)}_{b}} = {{(X-Y)}_{eff} - 0.7 G},
   \end{equation}
   
   \noindent where $(X-Y)_{eff}$ is the effective color index, measured within the effective 
   aperture in the B band. We have also define a characteristic total color
   as the one measured within 2 effective radius, corresponding therefore to
   
   \begin{equation}
   {{(X-Y)}_{T}} = {{(X-Y)}_{eff} + 0.3 G}.
   \end{equation}
   
   \noindent Equations (2) and (3), and the effective color indices given by the RC3,
   were used to determine these characteristic colors.
   
   As we have already mentioned, this second method is more suitable to compare the 
   values from different galaxies. However, we could not use this method to verify
   the correlation between the total colors and the colors of bulges, since Eq(s). 
   (2) and (3) already imposes such a correlation, as one can see by subtracting them.
   Therefore, this justifies our first rough method used only to verify the 
   existence of a real correlation, since that method do not suffer from this kind of bias.
   
   We have corrected these characteristic color indices for Galactic reddening using the maps
   of \cite{sch98} to obtain the (B\,$-$V) color excess, and used the relation
   
   \begin{equation}
   {{E(U-B)}\over{\;E(B-V)}} = {0.72\pm0.03},
   \end{equation}
   
   \noindent which can be found in \cite{kit98}.
   
   We did not correct these values for any differential internal reddening between bulge
   and disk. Instead, we have applied an integrated correction to account for the effects 
   of inclination. According to \cite{gio94}, 
   the internal extinction as a function of the inclination of the galaxy, 
   derived from I--band images of Sc galaxies, is
   
   \begin{equation}
   {{A}_{I}} = {1.12(\pm 0.05) \log {{a}\over {b}}},
   \end{equation}
   
   \noindent where $a$ and $b$ are, respectively, the major and minor axis of the galaxy. 
   For the U, B and V bands, \cite{elm98} shows that the extinction coefficients are, 
   respectively, 3.81, 3.17 and 2.38 times the extinction coefficient in the I band, 
   according to observations done in the Galaxy. Since the Galaxy is likely a Sbc galaxy
   we used these same relations for the objects in our sample. Using the definition
   of the color excess and the fact that $\log a/b$ is approximately 
   equivalent to the $\log R_{25}$ parameter of
   the RC3, we finally arrive to the relations used in our work:
      
   \begin{equation}
   {E(U-B)} = {0.68 \log R_{25}}
   \end{equation}
   
   \noindent and
   
   \begin{equation}
   {E(B-V)} = {0.87 \log R_{25}}.
   \end{equation}
   
   It is interesting to observe that the corrections we have applied are 
   actually $\sim$ 2--3 times larger
   than the ones adopted in the RC3! Indeed, earlier works (see, e.g., \cite{dev59}) argued
   that spiral galaxies were nearly transparent. 
   But more recent studies (e.g., \cite{bos94,gio95}) show that the optical 
   thickness of spiral galaxies is higher. The adopted corrections in Eq(s). (6) and (7) assume
   that spiral galaxies have a large optical thickness, and thus are much more realistic.
   
   Although the galaxies in our sample can be considered as local
   (-295 Km/s (NGC 224) $\leq cz \leq$ 
   8720 Km/s (UGC 4013)), with a typical value of $cz \sim$ 2000 Km/s), we 
   have also applied the K--correction, using the equations of the RC3.
   
   The galaxies analyzed in this work, as well as the results from the determination of the (B\,$-$V) 
   and (U$-$B) gradients, and of the total and bulge color indices, can be seen in Table 1.
       
\section{Comparative Studies}   
     
   \subsection{CCD Images}
   
   The ideal set of data to study the radial color distribution in the disk and bulge
   components is obtained by using CCD photometry, which permits a differential evaluation
   of the color along the galaxies. However, as mentioned before, we have choose a more fast way
   in order to have a statistically significant set of data. This was the main reason which lead us to use 
   the available data from LdV83,85. It is interesting, therefore, to compare the color distribution
   obtained from CCD and aperture photoelectric photometry.
   
   In this subsection, we present a comparison
   with the CCD data of 14 galaxies observed at the Pico dos Dias Observatory
   (PDO/LNA -- CNPq, Brazil). The CCD observations were done with a 24 inch telescope having a 
   focal ratio f/13.5, and using a thin back--illuminated CCD SITe SI003AB, with 1024 
   $\times$ 1024 pixels. The plate scale is 0.57 arcsec/pixel, resulting in a field
   of view of approximately 10 $\times$ 10 arcmin. The CCD gain 
   was fixed on 5 electrons/ADU and the read--out noise on 5.5 electrons.
   All objects were observed in the B, V, R and I passbands of the Cousins system.
   For each object, we have done 6 exposures in the B band, 5 in the V, and 3 in the R and I
   bands, typically, with an exposure of 300 seconds. The multiple
   exposures aim to ease cosmic ray removal. The data was calibrated with a set of standard stars
   of \cite{gra82} and corrected for atmosphere and Galactic extinction. The later correction
   was done using the maps of \cite{sch98}.
   
   The standard processing of the CCD data includes bias subtraction, flatfielding
   and cosmetics. The first step in the sky subtraction was done by editing the combined 
   images in each filter, removing the galaxy and stars. After that step we
   determined the mean sky background and its standard deviation ($\sigma$). Then, 
   we removed all pixels whose values were discrepant by more than 3 $\sigma$ from
   the mean background. An sky model was obtained by fitting a linear surface to the 
   image, and this model was subtracted from the combined image. We finally removed
   objects such as stars and H{\sc ii} regions. All these procedures were done
   using the {\sc iraf}\footnote{IRAF is distributed by the National Optical Astronomy Observatories,
   which are operated by the Association of Universities for Research
   in Astronomy, Inc., under cooperative agreement with the National
   Science Foundation.} package.
   
   We then used the {\sc ellipse} task to calculate the surface brightness profiles of each galaxy
   in each band. Subtracting the profiles we obtained color gradients, constructing tables in the
   same units of the ones in LdV83,85. These tables were used in the {\sc progress} algorithm to provide
   values for the gradients in the same way it was done for our whole sample.
   
   In Fig. 3, we show a plot of the CCD gradients and those obtained with the 
   photoelectric aperture data, showing that both estimations are essentially the
   same. The good correlation between these two set of values (Pearson correlation
   coefficient R = 0.93) gives support to the results obtained with the LdV83,85 data.
   The mean difference is G$_{LdV}$ $-$ G$_{CCD} \simeq -0.06$.
  
\placefigure{fig3}

   We have also done a comparison with the CCD observations made by \cite{dej94}
   to study color profiles in a sample of 86 face--on disk galaxies. We have 
   applied for the 8 galaxies our samples have in common, the same method we have used in
   this work, using the B and V CCD images kindly provided by de Jong. We have simulated 
   photometric apertures on these images using the {\sc imexamine} task from {\sc iraf}.
   The comparison of the (B\,$-$V) gradients obtained using the photoelectric data by 
   LdV83,85 and de Jong's CCD images revealed a Pearson correlation coefficient of $R=0.74$.
   If we do not consider two outliers the Pearson coefficient is $R=0.97$.
   
   \subsection{Comparison with Prugniel \& H\'eraudeau}

   \cite{pru98}, hereafter PH98, have also estimated the (U$-$B) and (B\,$-$V) gradients,
   using both CCD and photoelectric aperture photometry, for a large fraction of the galaxies 
   in our sample. To avoid the uncertainties due to inconsistent measures these authors have
   attributed different statistical weights to each source of data.
   
   In Fig. 4 we present a comparison between the gradients determined in the present work
   and those estimated by PH98. The correlation coefficient R is 0.85 for (B\,$-$V) 
   and 0.81 for (U$-$B). Moreover, we can see that there are no systematic differences 
   between the two works. The mean value of the differences
   is 0.004 in (B\,$-$V) and 0.011 in (U$-$B).
   
\placefigure{fig4}
   
\section{Analysing Gradients, Colors and Bars}

   In this section, we will analyse the results obtained from Sect. 3, regarding the 
   color gradients, as well as those relating to the total and bulge color indices. We have
   separated our sample into barred (SAB+SB) and non--barred (S+SA) galaxies, in order to test
   the bulge formation in the evolutionary and monolithic scenarios. Since the identification
   of bars is much more difficult in edge--on systems, and the effects of dust extinction are
   minimized in face--on galaxies, we took the special care of analysing the
   face--on and the edge--on galaxies separately. We use the same criterion as \cite{dej94},
   defining as face--on those galaxies with $\log R_{25} \leq 0.20$, 
   corresponding to $b/a \geq 0.625$. Galaxies which do not obey this criterion we
   regard as edge--on. The following galaxies, IC 983, NGC 253, NGC 1169, NGC 1625, NGC 1964,
   NGC 2276, NGC 2377, NGC 2525, NGC 3344, NGC 3646, NGC 4394, NGC 4402, NGC 5054, NGC 6215,
   NGC 6300, NGC 6878A, NGC 7307 and UGC 11555, whose gradients were too uncertain to be used,
   were removed from our sample.
   
   Table 1 shows the color gradients and its errors for all galaxies in our sample, as well
   as the bulge and total characteristic color indices. The errors of the gradients are
   the ones obtained through the {\sc progress} algorithm and thus are fit errors, which
   are larger than the photometric errors alone. One can see that the mean error on the
   (B\,$-$V) gradient is 0.03, and on the (U$-$B) is 0.05.
   The mean errors for the bulge and total color indices are, respectively, 
   0.04 and 0.03 for (B\,$-$V) and 0.05 and 0.04 for (U$-$B).
   
   \subsection{Gradients' Distributions}
   
   The distribution of the color gradients for barred and non--barred galaxies, considered 
   separately, both face and edge--on projections, can be seen in Fig. 5. The statistical data 
   from this figure are presented in Table 2, where column (1) contains the description of 
   each subsample, while columns (2) and (5) contain the total number of objects in each 
   subsample in each color index. Columns (3) and (6) show the mean values and their 
   respective standard errors. Finally, columns (4) and (7) contain the standard deviations 
   of these distributions. These values were obtained through a Gaussian fit to the observed 
   distribution.

\placefigure{fig5}

\placetable{tbl-2}
   
   We can observe from Fig. 5 that the (U$-$B) distribution for barred galaxies, for both the
   face and edge--on projections, is wider than the distribution for non--barred galaxies. 
   The results in Table 2
   also show that barred galaxies have wider distributions. From this table, we can
   see that the differences in the standard deviations are larger than the expected 
   photometric errors, indicating that this is indeed a real effect. With a
   smaller amplitude, the same effect is also present in the (B\,$-$V) gradients.
   Even considering
   that the photometric errors are larger in the U band, this can hardly explain
   this effect since such errors affect both kinds of objects, barred and non--barred, 
   in the same way. Therefore, this is a real characteristic of barred galaxies, namely, 
   to present a larger interval of (U$-$B) color gradients, probably associated with
   recents episodes of star formation. We note that the larger distributions are caused
   by a larger fraction of barred galaxies having zero or positive gradients.  
   In (U$-$B), for instance,
   55\% $\pm$ 8\% of the face--on barred galaxies have zero or positive gradients, whereas for the
   face--on non--barred galaxies this fraction is reduced to 32\% $\pm$ 12\%. However, considering
   the (B\,$-$V) index, these fractions are more similar, being 41\% $\pm$ 6\% among barred galaxies
   and 31\% $\pm$ 11\% among non--barred galaxies. At this point we might suspect that
   this difference between the two color indices may be caused by a larger 
   age/metallicity sensibility of the (U$-$B) color index. We remark that this effect
   is present even when we do not separate the edge and face--on galaxies.
   
   Moreover, one can see in Fig. 5 that the majority of barred galaxies have less pronounced 
   (U$-$B) gradients than the non--barred galaxies, as can be also verified through
   the mean values presented in Table 2. But, interestingly, this behaviour does not occur
   in the (B\,$-$V) color. This is probably an effect that any enhancement in the star formation
   rate affects more the (U$-$B) than the (B\,$-$V) color.
   
   Another interesting effect is that the edge--on galaxies show a tendency
   of having more pronounced negative gradients compared to the face--on systems, 
   specially in (U$-$B). This effect may well be related to the fact that 
   the internal reddening is more expressive in edge--on galaxies, and points
   to the presence of a small differential internal correction that affects the
   bulge and the disk in different ways. Indeed, one can conclude that the light 
   emitted by the central regions shall be more affected by reddening, a result
   that agrees with the ones presented by \cite{dej96c}.
   
   We present in Fig. 6 the (U$-$B) versus the (B\,$-$V) gradients for the non--barred
   galaxies (a), barred galaxies (b) and for the total sample (c). We can see from this
   figure that the gradients in both colors are well correlated, and that there is no 
   difference in the correlation for barred and non--barred galaxies.
   In fact, the Pearson correlation coefficient R is 0.71 for non--barred, 0.80
   for barred, and 0.78 for the whole sample.
   The same correlation was observed separating face and edge--on galaxies without
   noticeable differences. Again, we can see that barred galaxies have a more extended
   color gradient amplitude in these plots.
   These correlations are indeed expected, since the same physical
   reason rules the gradients in both colors, namely, variations between the stellar populations
   of the inner and outer regions of the galaxies. The models of \cite{lar78}, for instance, show
   that, for a population formed in a single burst, the variation in (B\,$-$V) for populations
   with a difference in age of 10 Giga years is 1.1, while for (U$-$B) it is 1.5.
   Thus, in these conditions, we shall expect $\Delta (U-B) / \Delta (B-V) = 1.4$. Since the
   color gradients are $G_{B-V} = \Delta (B-V) / \Delta \log {A}$ and $G_{U-B} = \Delta (U-B) / 
   \Delta \log {A}$, then we shall have $G_{U-B} / G_{B-V} = 1.4$. Surprisingly, the correlations
   in Fig. 6 give us $G_{U-B} / G_{B-V} = 1.2$, which is very close to what is predicted from these
   simple models. This difference might indicate that we are seeing stellar populations mixed
   with dust, since Larson and Tinsley's models do not take dust into account.
   
   Thus we interpret that the reason for this agreement, as will be seen in Sect. 5.5, is that the total
   color index is relatively stable among galaxies in our sample, but the color
   of the bulge varies noticeably between the barred and non--barred populations.
   Therefore, the amplitude of variation in the color gradients shown in Fig(s). 5 and
   6 is related to variations in the stellar population of the bulges.
   
\placefigure{fig6}   

   It is interesting to ask what would happen if the weakly--barred galaxies (SAB's) would
   have been analysed separately. The answer to this question is the analysis would remain
   the same. Indeed, barred and weakly--barred galaxies show essentially the same mean color 
   gradient in both the (B\,$-$V) and (U$-$B). The values for barred galaxies alone are
   $-0.12 \pm 0.02$ and $-0.11 \pm 0.03$, respectively, while for the weakly--barred galaxies these
   values turn to $-0.14 \pm 0.02$ and $-0.13 \pm 0.03$.
   
   \subsection{Negative, Zero and Positive Gradients}

   The vast majority of the galaxies in our sample have negative gradients, as one can see
   from Fig. 5, implying therefore that the bulge is redder than the disk. This result, in
   principle, is consistent with the monolithic scenario, where the older and redder population
   is located in the central parts, whereas the younger and bluer populations are
   more predominant in the outer regions of spiral galaxies.
   
   In order to get some further insight, we have considered three arbitrary categories for the color
   gradients, according to their values. The first category is constituted by objects
   having negative gradients, with $G \leq -0.10$; the second have galaxies with almost
   zero gradient, defined by $-0.10 < G < 0.10$, and finally the third category have those
   galaxies with positive gradients, $G \geq 0.10$. In Table 3 we show for the face--on galaxies
   in our sample, where the distinction between barred and non--barred is more reliable, 
   the distribution among these three classes of objects, in both colors. There are a total of 124
   face--on galaxies with the (B\,$-$V) gradient, and 104 with the (U$-$B). In column (1) we present
   the total number of galaxies in each class of color gradient, while column (2) gives their 
   fraction of the total sample. Column (3), (4) and (5) show, respectively, the fractions of 
   non--barred, weakly--barred and barred galaxies in each gradient interval. Column (6) shows
   the total fraction of barred (SAB+SB) galaxies and, finally, Column (7) shows the number
   of galaxies hosting AGN's. Galaxies with AGN were identified through the catalog of
   \cite{ver98}. The reason to investigate this class of galaxies in this 
   study comes from the suggestions presented by other authors 
   (e.g., \cite{shl89,shl90}) that bars can fuel AGN through processes similar to the ones of the 
   secular evolution. We can verify that, with small variations in each color index, we
   have approximately 59\% of the galaxies presenting negative gradients, 27\% with zero
   gradients, and 14\% with positive gradients. We remark that this result does 
   not change considerably when
   we consider a more restrictive definition of the zero gradient class, as $-0.05 < G < 0.05$. 
   Moreover, essentially the same result is also obtained when we consider the whole sample, including 
   together face and edge--on galaxies.

\placetable{tbl-3}   
   
   The total fraction of face--on barred galaxies in our sample is 79\%. We can see in Table 3
   that there exists an excess of barred galaxies among the ones with null or positive gradients. 
   In (B\,$-$V), 
   the fraction of barred galaxies with negative gradient is 75\%, while it raises to 91\% among
   the ones with zero gradient. In (U$-$B), 73\% of the galaxies with negative gradient are barred, 
   while 83\% of the ones with null gradient are barred, and 90\% of the positive gradient
   galaxies are barred. If we consider the more restrictive
   criterion for null gradient ($-0.05 < G < 0.05$), this excess is substantially emphasized.
   The fraction of barred galaxies with null (B\,$-$V) gradient then raises to 94\% and
   the fraction of barred galaxies with null (U$-$B) gradient raises to 88\%.
   This result indicates that barred galaxies are over--represented among the
   objects having null or positive gradients. Therefore, bars seem to act as a mechanism
   of homogenization of the color indices, and thus, of the stellar population, 
   along galaxies. As a consequence, we are forced to conclude that a classical 
   monolithic scenario would have difficulties to explain this result.
   
   Another interesting feature of Table 3 is that the fraction of
   galaxies with AGN increase from $\sim$ 8\% for systems with negative gradients to 
   $\sim$ 36\% for objects with positive gradients. Even considering the low number
   statistics, this might be an indication that the homogenization of the stellar population, 
   induced by bars, is related to the AGN phenomenon.
   
   \subsection{Color Gradients and Abundances}
   
   Recent theoretical studies (e.g., \cite{fri95}) related
   to the dynamical secular evolution show that a stellar bar is able to collect gas from the 
   outer to the inner regions of the disk, through shocks and gravitational torques that remove 
   angular momentum from the gas. Thus, a large--scale mixing of the gas must occur along the galaxy, 
   which could be, in principle, observed in the radial abundance profiles of certain chemical 
   elements.
   \cite{mar94}, hereafter MR94, and \cite{zar94}, hereafter ZKH94, present
   O/H abundance gradients in spiral galaxies determined through the observation of H{\sc ii}
   regions. Both studies show that barred galaxies tend to have less pronounced gradients. 
   Moreover, MR94 conclude that the gradients become less pronounced as the
   normalized length of the bar, or its apparent ellipticity, increases.
   On the other hand, studies from \cite{sak99} show that barred galaxies have a higher 
   central concentration of molecular gas (CO) than non--barred galaxies. Both results 
   are in agreement with the prediction from the theoretical studies of dynamical secular 
   evolution. Then, if the abundance is affected by this mechanism we also should expect it 
   affected the color gradients.
   
   In order to verify this possibility, we have compared 12 galaxies in common with MR94, and 18 with ZKH94.
   In Fig. 7, we plot our color gradients versus the abundance gradients of MR94 (top panel) and ZKH94
   (bottom panel). We can see that there is no clear correlation between the photometric and
   the abundance gradients. Hardly this absence of correlation could be a consequence of errors in the 
   photometric gradients, which typically range from 0.02 to 0.05. 
   On the other hand, the errors in the abundance gradients are more difficult to determine, 
   as we can see by looking at the quite different values of the NGC 2997
   gradient as estimated by MR94 and ZKH94. However, these
   errors are also hardly larger than 0.02 dex $\times$ kpc$^{-1}$ (ZKH94).    
   One can interpret that the
   absence of such correlation is a real feature, and thus it is interesting to explore its consequences.
   Since the color indices are sensible to both age and metallicity, this result could indicate
   that the excess of barred galaxies
   with zero color gradients, as we found in Sect. 5.2, reflects a difference in 
   the behaviour of the mean {\em age} of the stellar population along 
   barred and unbarred galaxies, and not of its metallicity.
   However, in principle, this absence of correlation could be attributed to the effects of dust extinction.
   We have argued in Sect. 3.1 that these effects shall be small, but in Sect. 5.6 below
   we will show a quantitative analysis of these effects, and we conclude that it is possible that the lack
   of this correlation may be caused by dust extinction.

   We would expect to find such correlation in the dynamical secular evolutionary scenario.
   However, if we consider that bars are a relatively fast recurrent phenomenon, this absence of
   correlation would be natural. Indeed, we can imagine the following picture. If we consider
   a galaxy formed through the monolithic scenario, we shall expect it to show both the abundance
   and color gradients negative. In that case the galaxy would be placed in the lower left region of
   Fig. 7. This galaxy can develop a bar and then have its abundance gradient
   shallower, while its color gradient shall remain the same, because the time scale to mix the
   gas in the disk shall be smaller than the time required to form new stars in the central region.
   Galaxies in that
   stage would occupy the lower right part of Fig. 7. After the gas accumulates in the central
   region it will form new stars and then the color gradient will become shallower and the galaxy
   would be in the upper right part of Fig. 7. Instabilities generated by the mass accumulated in the
   central region will destroy the bar interrupting the transfer of gas along it and steepening the
   abundance gradient, while keeping the color gradient unchanged. In this case, we will see the galaxy
   in the upper left part of Fig. 7. The lack of new star formation in the
   central region and the aging of the stars will then turn negative the color gradient and the
   galaxy will again occupy the lower left part of Fig. 7. If a new bar is developed then the
   changes in the abundance and color gradients can occur again.
  
\placefigure{fig7}
   
   \subsection{Color Gradients and the Morphology of Bars}
   
   In an attempt to perform a quantitative morphology of bars in galaxies, \cite{mar95}, 
   hereafter M95, made visual estimates of the axial ratio, $b/a$, the major axis length (normalized
   by the 25 mag arcsec$^{-2}$ isophote), $L_{b}$, and the 
   apparent ellipticity of bars in spiral galaxies.
   In that work, it is found a relation between the length of the bar and the diameter of the bulge, 
   in the sense that galaxies with large bulges also have large bars. Moreover, he found an
   apparent correlation between the presence of intense nuclear star formation and the axial
   ratio of the bar, in the sense that strong bars, those with $b/a \leq 0.6$, are
   present in galaxies with nuclear bursts of star formation.
   
   A total of 45 galaxies in our sample were studied in M95, allowing us to verify 
   correlations between our color gradients and the parameters of the bar morphology. Figure 8
   shows for these objects our color gradients plotted against the bar parameters 
   axial ratio, $b/a$, length, $L_{b}$, and apparent ellipticity, $\varepsilon_{b}$.
   We detect no correlation of these morphological bar parameters with the color gradients, 
   meaning that the color gradient does not depend on the morphology of the bar. 
   Thus, there are galaxies with the same gradient and bars with quite distinct morphologies.
   And, on the other hand, there are systems with the same bar morphology and quite 
   different color gradients. It is worth notice that MR94 found that the O/H abundance
   gradients in barred galaxies turn less pronounced as the ellipticity or the length of
   the bar increases, i.e., galaxies with stronger bars have less pronounced O/H
   abundance gradients. Again, it is not unlikely that extinction by dust is masking a
   correlation. However, these results may be explained by different time scales in the
   homogenization of abundance gradients, measured in gas, and color gradients,
   measured in stars.
  
\placefigure{fig8}

   \subsection{Total and Bulge Color Indices}
   
   We remark that our total color indices are obviously affected by the contributions of
   both the bulge and the disk. The relative importance of these two components can be
   measured by the factor $f_B = L_{Bb}/L_{Bd}$, representing the 
   bulge to disk luminosity ratio in the B band.
   On the other hand, both components have intrinsic colors (B$-$V)$_d$, (B$-$V)$_b$ and
   (U$-$B)$_d$, (U$-$B)$_b$. The total color is related to these component colors 
   through the relations

   \begin{equation}
   (B-V)_T = (B-V)_b - 2.5 \log\frac{f_B+1}{\;\;\;\;f_B + 10^{ 0.4\Delta_{BV}}}
   \end{equation}  

   \noindent and

   \begin{equation}
   (U-B)_T = (U-B)_b + 2.5 \log\frac{f_B+1}{\;\;\;\;f_B + 10^{-0.4\Delta_{UB}}}
   \end{equation}  

   \noindent where $\Delta_{BV}=(B-V)_d-(B-V)_b$ and $\Delta_{UB}=(U-B)_d-(U-B)_b$.

   Table 4 shows the median values of the characteristic total and bulge color indices for
   the galaxies in our sample, separated by the gradient class, together with their standard
   errors. For those objects with null
   color gradient we show a single color value. In the right part of this table, we
   present the data relative only to the face--on objects. We can see that the same trend
   is present in both samples.

\placetable{tbl-4}   
   
   Considering both the face--on galaxies and the total sample, one can observe that the 
   total colors remain almost with the same value for the three classes of gradients.
   The differences are small in both the (U$-$B) and (B\,$-$V) colors, within the errors.
   However, bulges of zero or positive gradient objects are systematically bluer than the ones found
   in negative gradient objects. The differences are much larger than the errors, 
   indicating that it is a real effect. Indeed, there is a difference of order 0.40 magnitudes
   between the colors of bulges in negative and positive gradient objects, while the 
   estimated errors are within $\sim$ 0.03 magnitudes. Therefore, one major factor
   determining the value of the gradient is the bulge color. Moreover, the disk colors should
   also be redder, for objects with null or positive gradients, in order to keep the total
   colors almost unchanged, as it is observed. This is an effect which is not compatible
   with the monolithic scenario, since it indicates that, in the process of homogenization
   of the stellar population, induced by bars, bursts of star formation occur in the bulge, 
   in complete agreement with the secular evolutionary scenario.
   
   Another way of looking to this effect is shown in Fig. 9, where we plot the relation of the total 
   and bulge color indices for different classes of gradients, considering only face--on galaxies.
   Although we have the total color 
   instead of the disk color, these correlations have the same meaning as the ones found by 
   other authors (\cite{pel96}), showing that the formations of bulge and disk are parts of the
   same process. However, this figure also shows that the zero point scale of the correlation
   is quite different for objects having negative and positive color gradients. While the 
   correlations are in the same sense, we can see again that the bulge is much bluer in objects
   with positive gradients, while the mean total color is the same, irrespective of the
   gradient category. These results do not change when we consider also the edge--on galaxies.
   
\placefigure{fig9}

   Once again, it is interesting to verify if there are any differences in the properties
   of barred and weakly--barred galaxies. Like the color gradients, the characteristic total 
   and bulge mean color indices for SB's and SAB's are essentially the same. The bulge colors for
   SB's are $0.56 \pm 0.02$ and $-0.01 \pm 0.04$ in (B\,$-$V) and (U$-$B), respectively, while for
   SAB's they are $0.60 \pm 0.02$ and $0.06 \pm 0.03$. On the other hand, the total colors for
   SB's are $0.45 \pm 0.02$ and $-0.11 \pm 0.02$ in (B\,$-$V) and (U$-$B), and 
   they are $0.46 \pm 0.02$ and $-0.08 \pm 0.02$ for SAB's.
   
   \subsection{Dust Extinction}

   A fundamental point to be considered in this study are the effects of dust extinction and reddening.
   In principle, dust can disturb the analysis of color distribution in galaxies.
   To minimize its effects we have made a careful sample selection excluding galaxies presenting
   strong dust lanes.

   Moreover, we have made our analysis considering also a sub--sample
   containing only the face--on galaxies of our total sample, in which it is well known that
   the effects of dust are minimized. We also consider the results from the models
   of dust distribution in disk dominated
   galaxies by \cite{dej96c} which show that the dust reddening plays a minor role in color gradients.
   This author also argues that color gradients produced by dust are small from
   the U to the R bands because the absorption properties do not change very much in these bands.
   Furthermore, we have shown that there is an excess of barred galaxies with blue bulges in comparison
   with non--barred galaxies, and we conclude that this is related to recent bursts of
   star formation. Since the effects of dust do not depend on whether or not the
   galaxy hosts a bar, this main conclusion remains unaltered, even if the extinction is considerable.

   Nevertheless, although the extinction in face--on galaxies is
   smaller than in edge--on galaxies, it might be
   considerable in the central regions (see \cite{pel95}). Moreover, extinction and reddening depend
   on the geometry of the system and on the distribution of dust and stars (see, e.g., \cite{jan94}), so
   that it is prudent to verify empirically the role of dust in color gradients. With this aim, we have used
   HST archival data (NICMOS and WFPC2), and some CCD images obtained at Pico dos Dias, in order to determine
   the optical (B,V,I) and near--IR (H,K) color gradients
   for some galaxies, which are useful to evaluate the role of dust.
   These galaxies were chosen to have an inclination representative of our sample.
   As we have no photometry data in all selected passbands for all galaxies used in
   this analysis (see Table 5), we will assume that such gradients like ($H-K$) indicate variations
   in the old stellar population, while those like ($B-V$) or ($B-I$) are specially sensitive to
   recent star formation. Color gradients like ($I-H$) or ($V-H$) will primarily show the extinction
   caused by dust, as well as old stellar population gradients (see \cite{pel99}). All galaxies
   belong to our main sample (Sect. 2). As the HST data were measured only in
   the central region of the galaxies (inner $\sim$ 2 kpc), these central
   gradients shall not be compared with the global ones obtained in Sect. 3.

\placetable{tbl-5}

   Since it is in the central region where most of the dust is accumulated,
   its role in color gradients evaluated here may be considered as an upper limit.
   Let us evaluate firstly the HST data. As the dust and gas contribution
   are not the same for all galaxies, we will discuss the results for each one individually, and
   summarize them in Table 5. NGC 3310 shows a very small old population gradient
   ($G(H-K)=-0.04$) and a small old population/dust gradient ($G(I-K)=-0.11$),
   while the color gradient produced by recent
   star formation is large ($G(B-I)=-0.41$). Thus one can conclude that, for this galaxy,
   dust may be responsible
   for $\sim 17\%$ of the observed central color gradient. NGC 5033 have also a very small old population
   gradient ($G(V-H)=+0.01$) but a {\em positive} and large star formation gradient ($G(B-H)=+0.39$).
   This means that,
   even with the dust present in the centre of this galaxy (as can be seen in the HST images), the
   blue light emitted by the young population are strong enough to produce positive color gradients.
   Another possibility to explain this behaviour is the presence of a strong dust lane off--centered,
   but this lane was not found in the images. NGC 5194 also have a very small old population
   gradient ($G(H-K)=-0.02$) and a considerable old population/dust gradient ($G(V-K)=-0.21$).
   With the star formation
   gradient values one can conclude that, in this galaxy, dust may cause nearly half of the observed
   central color gradient. Finally, NGC 5248 have a considerable
   old population/dust gradient ($G(V-H)=-0.24$), but a {\em positive} star formation gradient. Conclusions
   are the same as for NGC 5033.

   Another way to study the role of dust in color gradients is to determine the reddening it causes.
   Using the HST data again we can estimate an upper limit, considering that there is no dust reddening
   beyond 1 $R_{eff}$ and that there are no stellar population gradients. Thus, the difference in color
   from the center to 1 $R_{eff}$ can be assumed to be all done by dust extinction. When the data
   does not reach 1 $R_{eff}$ we used the farthest available radius.
   We thus estimated such color excesses
   in ($I-K$) for NGC 3310, ($V-K$) for NGC 5194, and ($V-H$) and ($I-H$) for NGC 5248. Results are
   in Table 5. With the Galactic extinction law \cite{rie85} we have determined the extinction $A_{V}$
   in the centre of these galaxies. Its average value is $A_{V}=1.5$. \cite{pel99} applied the
   same analysis to a sample of early--type spirals, obtaining $A_{V}=0.6-1.0$.

   The same procedure we have used to the HST data we applied for 5 galaxies observed
   by us at the Pico dos Dias observatory in the
   B, V and I bands (see Table 5). Assuming that ($B-V$) gradients are sensitive to recent star
   formation, while ($V-I$) gradients are old population/dust gradients, we can infer the dust
   contribution to the observed color gradients to be of up to 45\% in the central region.
   As one can see, there are 2 galaxies
   with a negative old population/dust gradients but with positive star formation gradients.
   This result is in agreement with the one obtained using the HST data.
   We have also estimated an average value for $A_{V}$ using the ($V-I$) color excesses. Its value
   is $A_{V}=0.4$. This value is lower than the one obtained with the HST data simply because it
   was not obtained with optical--near--infrared colors.

   Now, assuming that the color excesses obtained truly represent an effect of dust extinction, we
   can ``correct'' the colors inside 1 $R_{eff}$ and re--calculate the ($B-V$) and ($U-B$) color
   gradients, using the Galactic extinction law. Table 6 shows the results and compares them with
   the gradients determined in Sect. 3. It can be seen that, with the HST data, dust effects can,
   in some cases, alter significantly the color gradients determined. But in other cases, even the
   high values of the upper limit for $A_{V}$ do not change the results. Table 6 also shows that
   using the color excesses obtained through our B, V and I CCD imaging make no significant changes
   in the color gradients.

   This study has led us to conclude that indeed extinction in the center (inner $\sim$ 2 kpc)
   of late--type spirals is
   high, with a typical value for $A_{V}=1-2$ magnitudes. However, the results shown here seems to
   indicate that dust is very much concentrated in the center, so that {\em global} color gradients
   are not much disturbed by dust, in general. The fact that, even with dust present in the center,
   some galaxies have positive gradients, shows that the excess of barred galaxies with blue
   bulges, found in this work, is a result which is not affected by our ignorance on the dust effects.
   It means also that, in these blue bulges, one can have an underlying old stellar population
   beneath a recent burst of star formation. On the other hand, it seems that the absence of
   correlations between color gradients and abundance gradients (Sect. 5.3), and color gradients
   and the bar morphology (Sect. 5.4), could possibly be explained by dust extinction.

\placetable{tbl-6}

\section{General Discussion and Conclusions}

   In the previous section we noticed that barred galaxies have less pronounced (U$-$B)
   mean color gradients. Moreover, both at (U$-$B) and (B\,$-$V) the amplitude of
   variation of the gradient, as measured by the standard deviation of its distribution, 
   is larger in barred, as opposed to non--barred galaxies. These results imply that there
   is an excess of barred galaxies among the objects with null or positive gradients, as
   can be seen from Table 3. As a consequence, we conclude that bars act in the sense of
   promoting a more homogeneous stellar population in late--type spirals. Besides an
   underlying old and red stellar population, disks of late--type spirals have
   ubiquitous young and blue stars. Bulges in general have an old stellar population, but
   we have shown here that bulges of late--type barred galaxies have also an important
   young stellar component. Therefore, the
   stellar population of barred galaxies tend to show a degree of mixing not compatible with 
   the pure monolithic scenario.
   
   We found no correlation between the color and abundance gradients.
   We must consider here the results of Sect. 5.6, i.e., dust extinction is considerable
   in the central region of late--type spirals, but does not strongly disturb global color
   gradients, in general. In spite of the caveat
   that this lack of correlation may be caused by the effects of dust, judging from the
   estimated photometric and abundance errors, we believe that this could be a real effect,
   indicating that color gradients may be not associated with metallicities
   (but see \cite{pel99}). Therefore, the
   presence of color variations inside a given galaxy is quite probably related to an
   age effect caused by bursts of star formation. The absence of this correlation
   could also be explained if we consider that bars are a fast recurrent phenomenon.
   
   Another conclusion from this study is that the mean total color indices remain
   remarkably constant independently of the galaxy's color gradient. From the sample
   of face--on objects we can verify in Table 4 that the total mean colors are
   (B\,$-$V)$_{T} \simeq 0.55 \pm 0.02$ and (U$-$B)$_{T} \simeq -0.02 \pm 0.06$.
   On the other hand, bulges behave quite differently. The mean colors of bulges in
   null gradient galaxies are $\sim$ 0.20 bluer than the colors of bulges in negative 
   gradient systems. Bulges of positive gradient galaxies are even bluer, $\sim$ 0.50
   bluer than bulges in negative gradient objects. We see also in Table 4 that this 
   difference is quite too large to be explained by photometric errors. In order to keep
   the total color unchanged it is necessary that the disks of the null or positive
   gradient galaxies become redder, i.e., evolve passively.
   
   This same effect can be clearly seen from Fig. 9, where we present the correlation 
   between total and bulge colors. In both the negative and positive gradient regimes
   there is a correlation between these two colors. These correlations are in 
   agreement with the ones found by other authors (\cite{pel96}) for the colors of 
   bulges and disks. However, we can also see from Fig. 9 that the correlation of the
   positive gradient objects is shifted in the blue direction by $\sim$ 0.50 magnitudes
   in their bulges. According to these authors, assuming similar metallicities
   for bulges and disks, their correlations imply in a difference of the order of less than 30\%
   between the ages of the stellar populations in these two components. Again, the presence
   of a correlation between the total and bulge colors, as well as the bluer colors of
   bulges in galaxies with null or positive gradients, are not consistent with the pure monolithic
   scenario.
   
   A more difficult task is to identify the correct evolutionary scenario responsible
   for these observable properties. The capture of nearby dwarfs in the accretion process
   of the hierarchical scenario seems to be incompatible with the constancy of the mean total
   colors of galaxies presenting different classes of color gradients, since this process
   do not predict a passive evolution for the disk. Moreover, the hierarchical scenario
   also does not predict an excess of barred galaxies showing null or positive color
   gradients.
   
   On the other hand, the secular evolution induced by a bar can result in an enhancement
   of the star formation rate in the central regions of galaxies. This effect can be
   responsible for the bluer colors observed in bulges of galaxies showing null or
   positive color gradients. At this point we can not say, however, whether this enhancement
   is occurring in the bulge or in the internal region of the disk.
   
\begin{acknowledgements}

   It is a pleasure to thank Ronaldo E. de Souza and Rob Kennicutt for fruitful
   discussions and suggestions, and for a careful reading of a preliminary version of the paper,
   and Tim Beers for presenting us the LMS method. 
   Special thanks go to Roelof de Jong for providing us his CCD observations.
   We also thank G. Longo for helpful answers to our questions.
   We thank the anonymous referee for helping to improve the article, specially the
   discussion on dust effects.
   We acknowledge the Conselho Nacional de Pesquisa e Desenvolvimento (CNPq), the NExGal -- ProNEx
   and the Funda\c c\~ao de Amparo \`a Pesquisa do Estado de 
   S\~ao Paulo (FAPESP) for the financial support. We would also like to thank the staff at the Pico
   dos Dias Observatory (OPD/LNA -- CNPq) for helping during the observational runs.

\end{acknowledgements}

\clearpage

\placetable{tbl-1}

\clearpage

\figcaption[exegrad.pap.eps]{Examples of color gradients. Filled boxes indicate the (B\,$-$V), 
while circles indicate the (U$-$B). The color indices in magnitudes are plotted
against the decimal logarithm of the aperture in units of 0.1 arcminute. Dashed lines 
refer to the LS method, while dotted lines refer to the LMS method and the solid lines refer 
to the {\sc progress} algorithm. \label{fig1}}

\figcaption[dust.pap.eps]{Color gradients for the galaxies in our sample plotted against
the color excess ($E(B-V)$) caused by Galactic reddening (top panels), and against
the inclination of the galaxy ($\log R_{25}$ parameter of RC3 -- bottom panels -- 
the most edge--on galaxies are at the right side). 
The absence of correlations
show that we can neglect the effects of Galactic and internal reddening when calculating
color gradients in galaxies. Typical error bars are drawn at the bottom--right of each
panel. \label{fig2}}

\figcaption[fig13.pap.eps]{Comparison between color gradients determined through photoelectric
aperture photometry data (LdV) and through surface photometry (CCD). The good correlation shown
atest the validity of the results obtained here. \label{fig3}}

\figcaption[prugniel-1.pap.eps]{Comparison between color gradients from PH98 and
from this work. The dashed line indicates a 
one--by--one correlation. The solid line indicates a linear fit. \label{fig4}}

\figcaption[dgrade-2.pap.eps]{Distribution of the (B\,$-$V) and (U$-$B) color gradients
determined for the galaxies in our sample. (a): non--barred face--on galaxies; 
(b): barred face--on galaxies; (c): non--barred edge--on galaxies and (d): barred edge--on galaxies. 
\label{fig5}}

\figcaption[dgrade-1.pap.eps]{The (U$-$B) gradients plotted against the (B\,$-$V) gradients for: 
(a) non--barred galaxies, (b): barred galaxies and (c): all sample. The straight line corresponds
to a linear fit. \label{fig6}}

\figcaption[martin-2.pap.eps]{Correlation between our photometric gradients and the O/H
abundance gradients of MR94 (top panel) and ZKH94 (bottom panel). Boxes indicate the
(B\,$-$V) gradient, while circles indicate the (U$-$B). Filled symbols refer to barred
galaxies. In the bottom right part of each panel typical error bars are shown. \label{fig7}}

\figcaption[martin-1.pap.eps]{Color gradients plotted against the morphological parameters of
the bars of M95. In the top panels the gradients are plotted against the axial ratio.
In the middle panels they are plotted against the normalized length of the bars, while
in the bottom panels a correlation with the apparent ellipticity was sought. \label{fig8}}

\figcaption[buldis1.pap.eps]{The correlation for the total and bulge characteristic color indices,
calculated through our first method, for galaxies with negative color gradients 
(upper panels) and positive color gradients (lower panels). The solid lines are fits
to the data points. The fits of the upper panels are also shown in the lower panels
(dotted lines) for comparison. Only face--on galaxies are considered. \label{fig9}}

\plotone{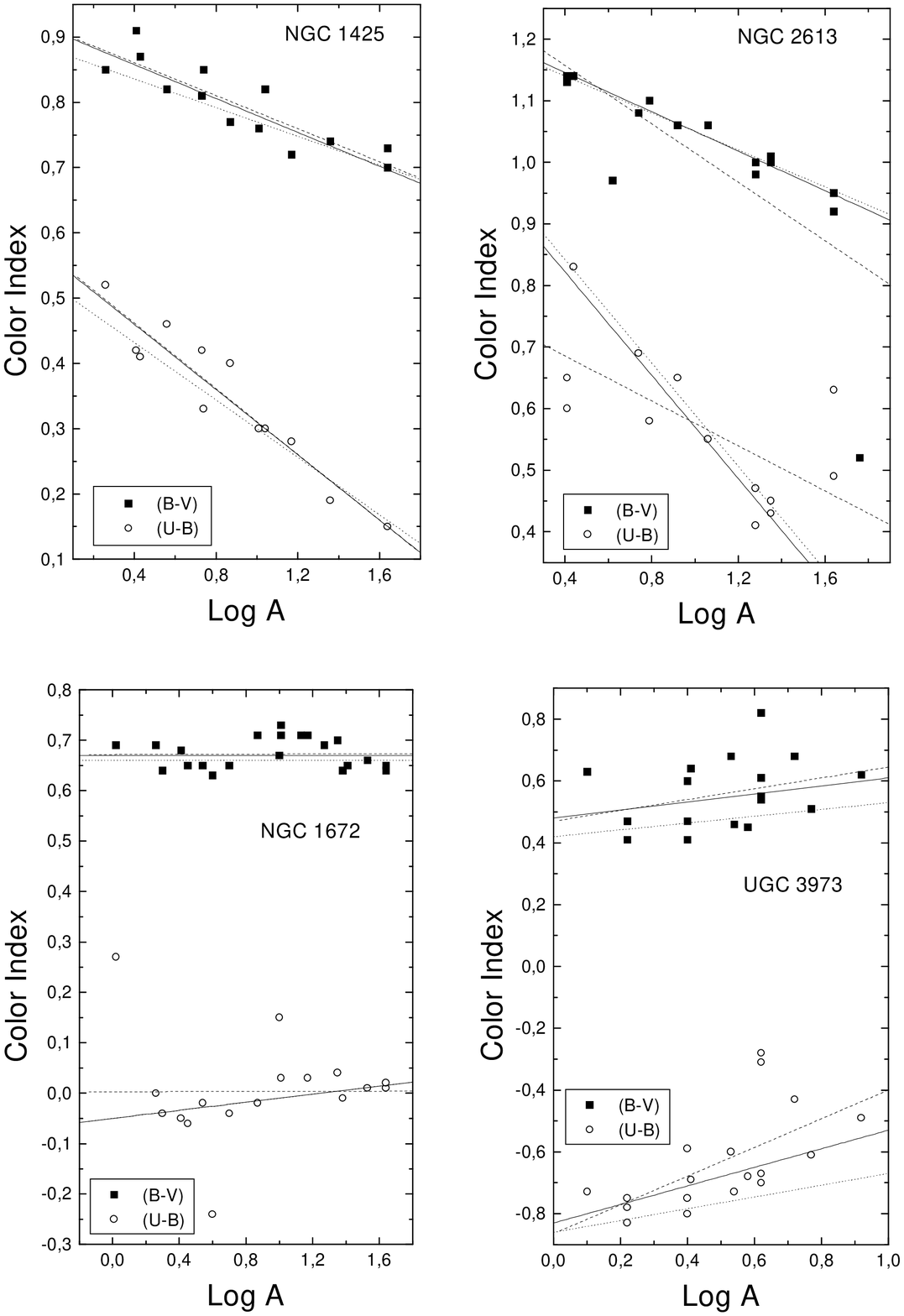}

\plotone{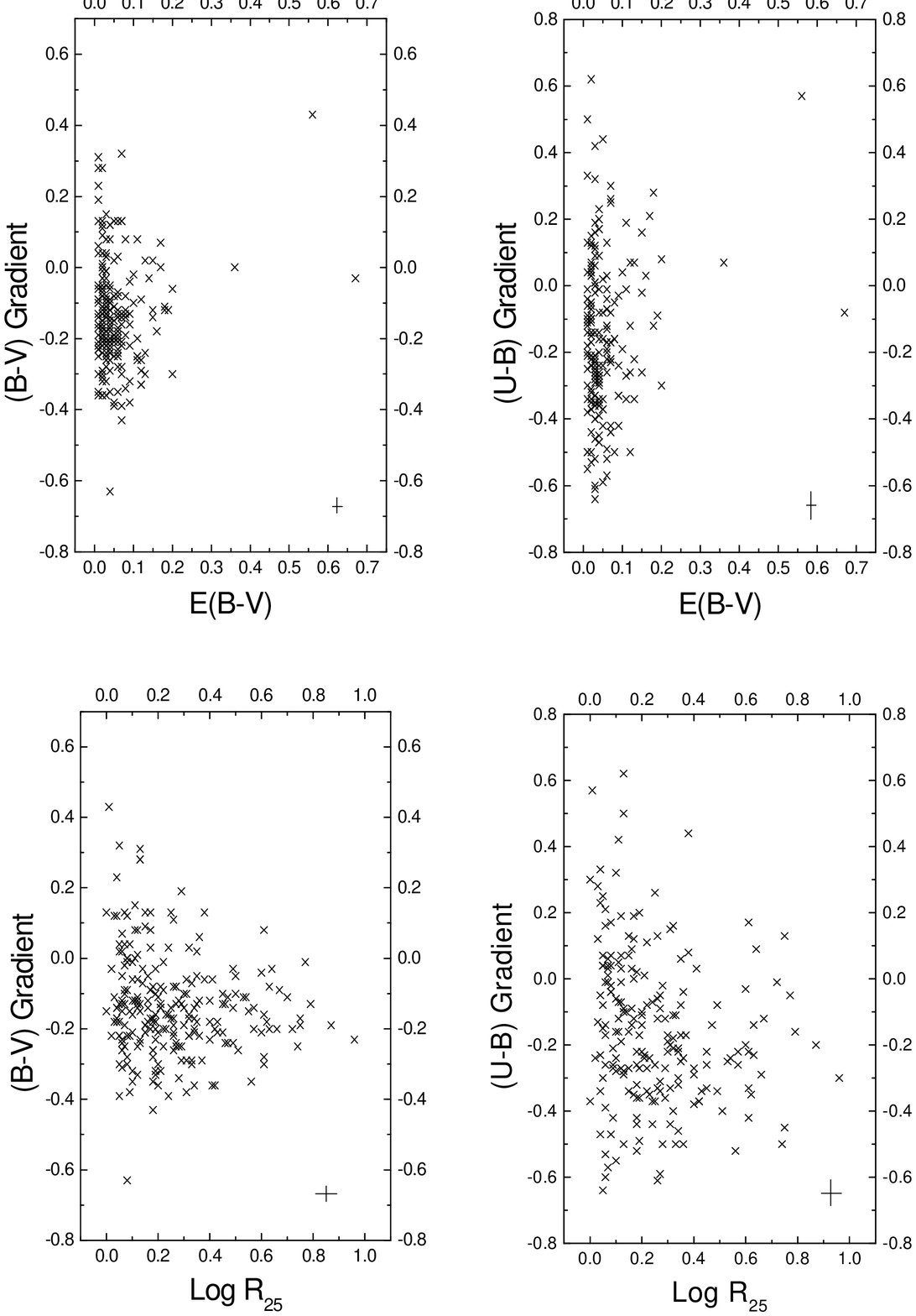}

\plotone{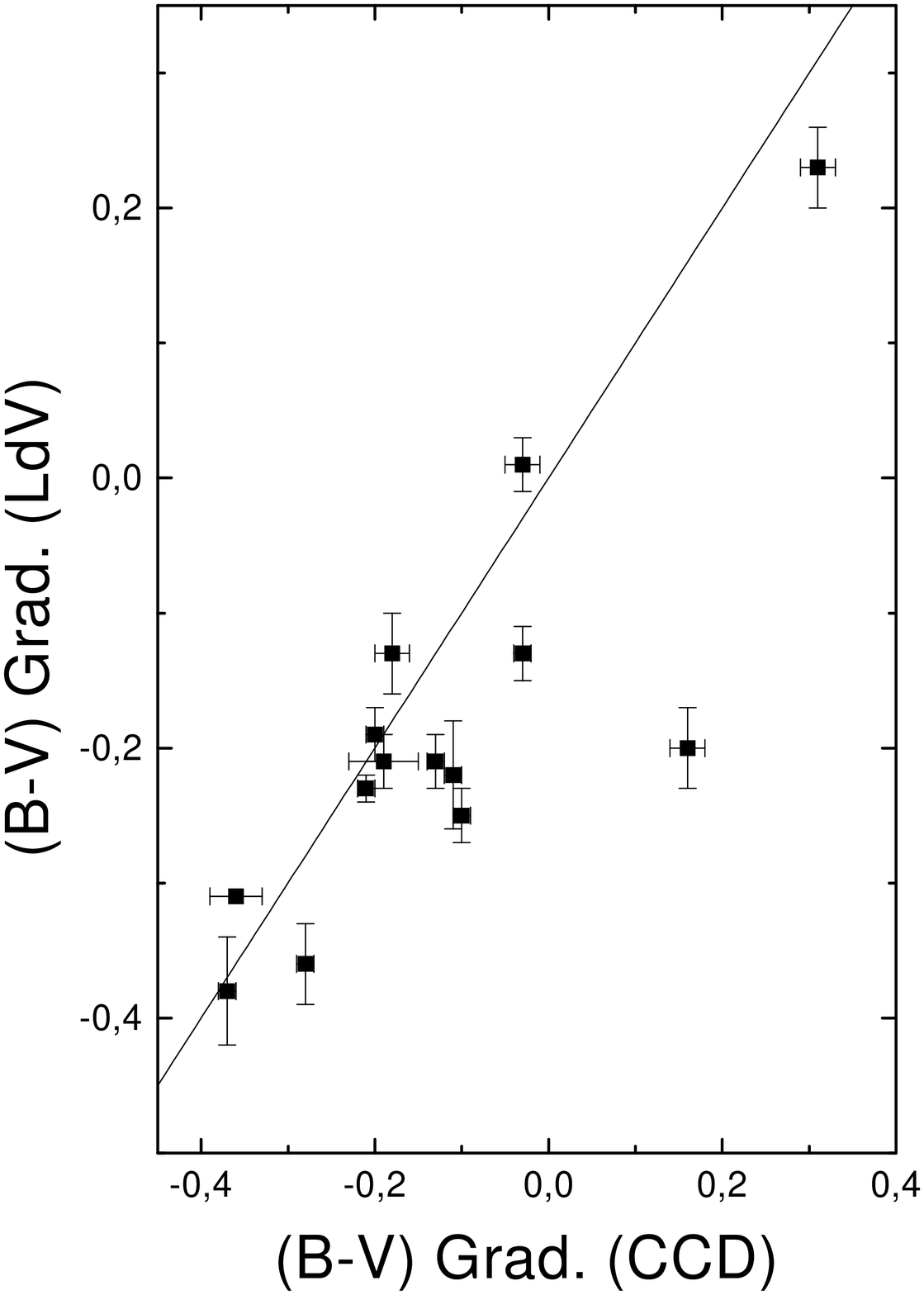}

\plotone{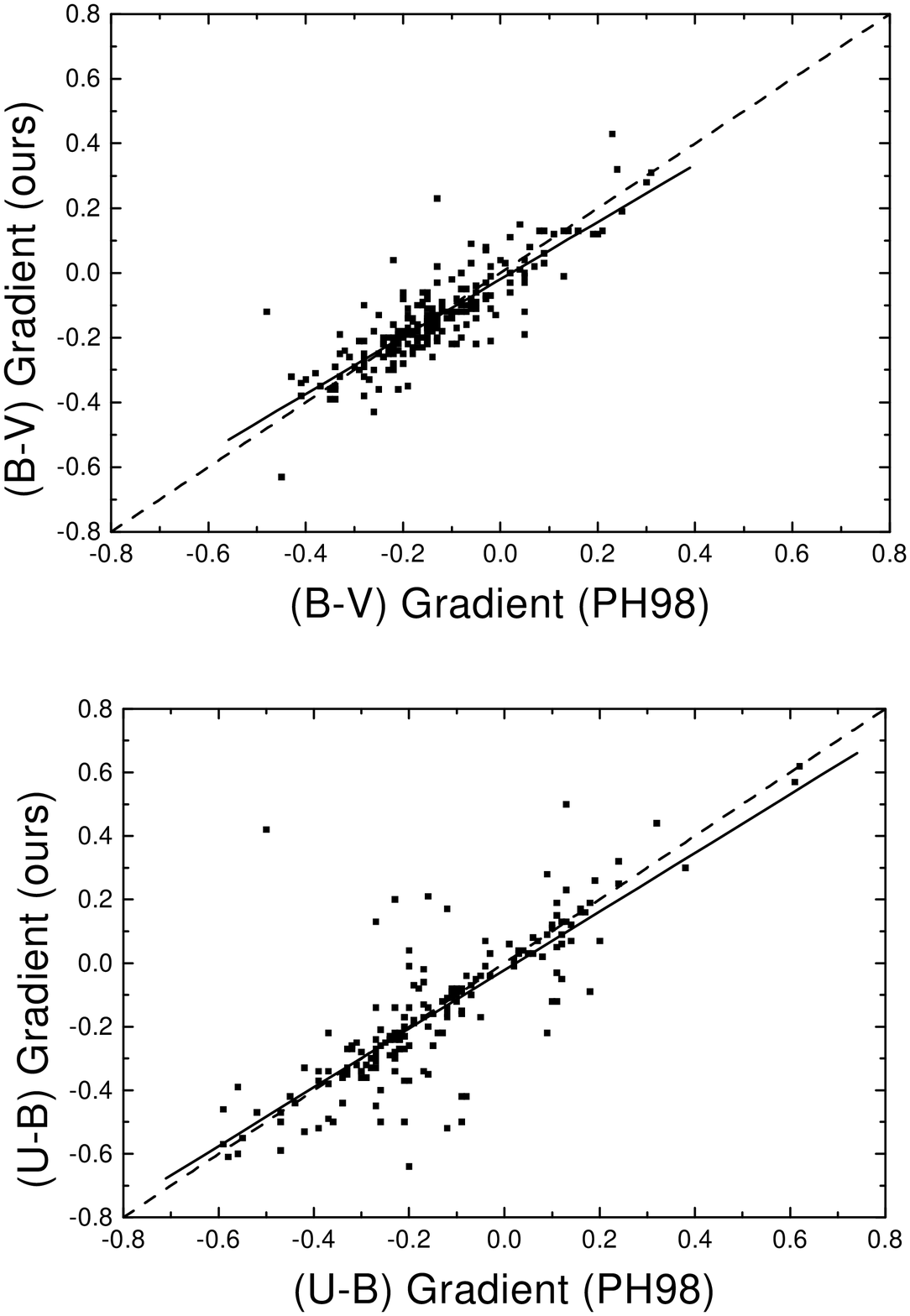}

\plotone{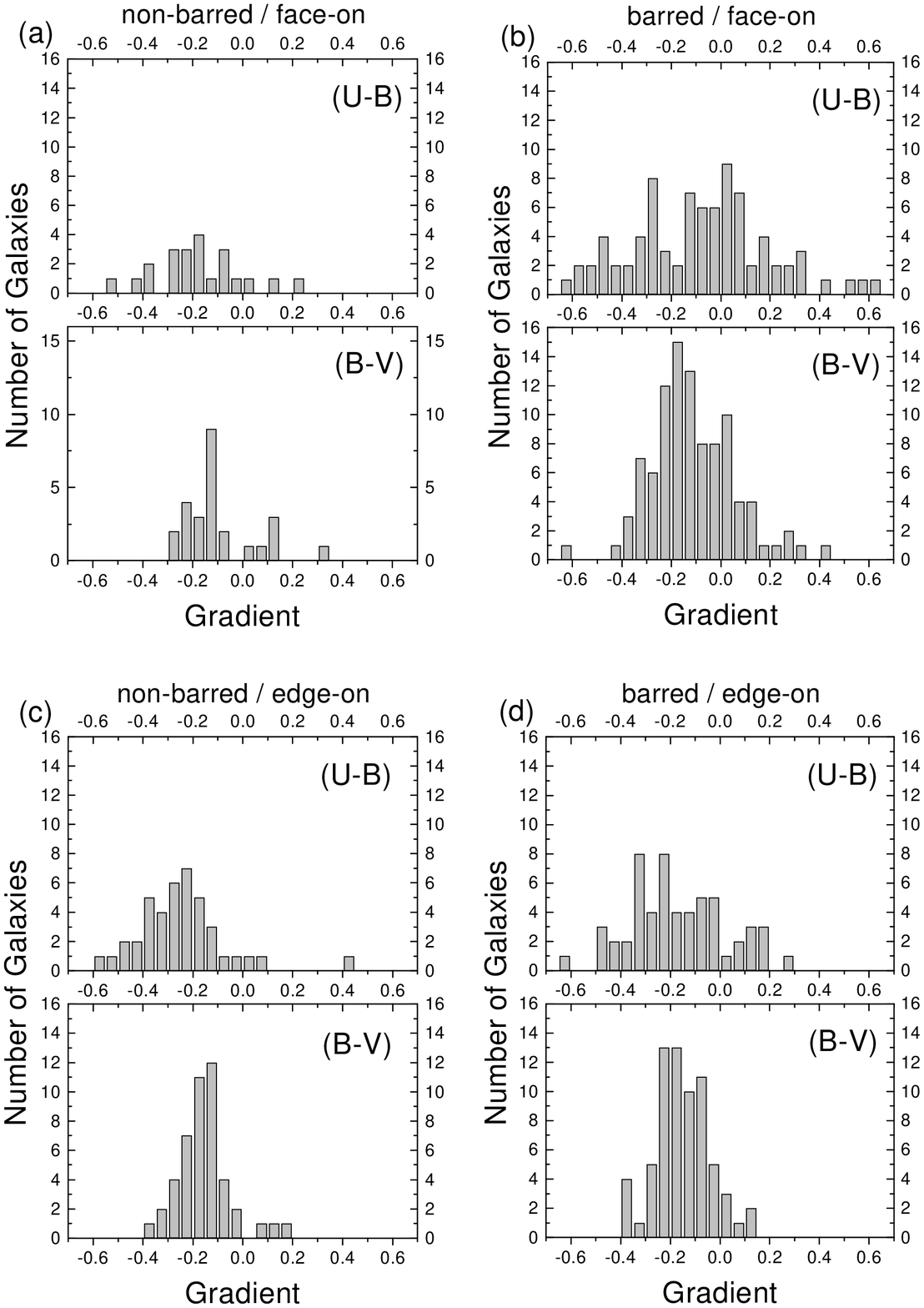}

\plotone{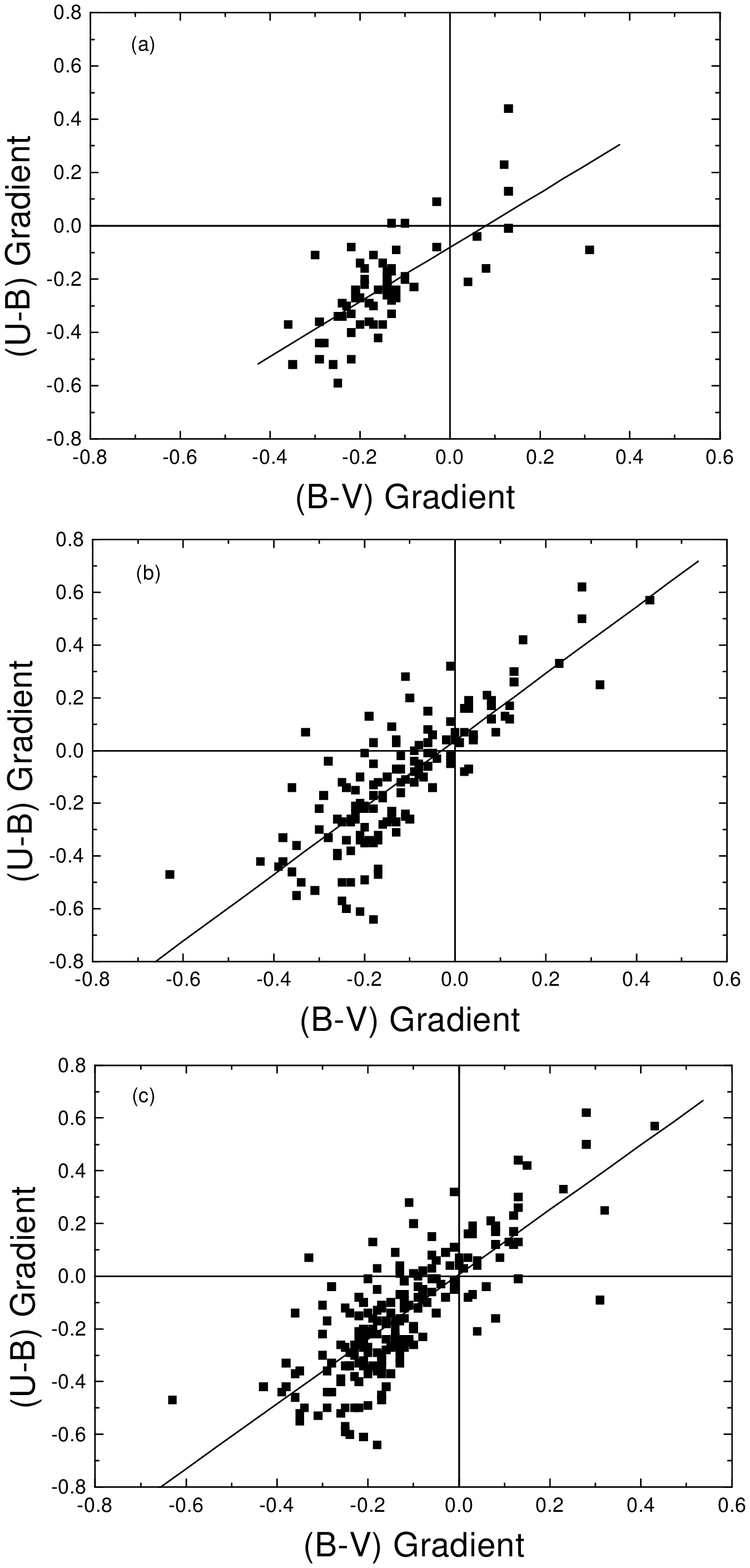}

\plotone{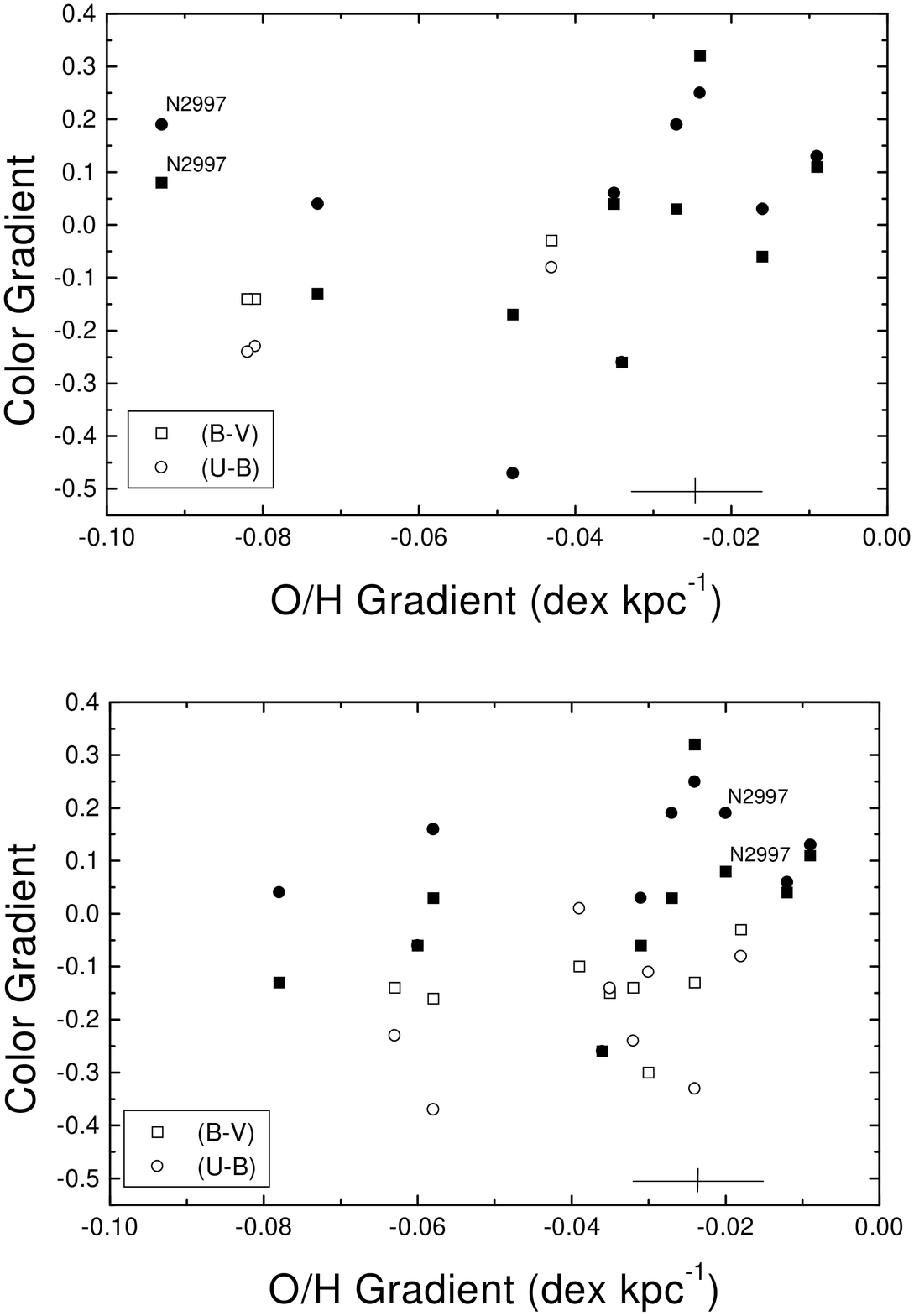}

\plotone{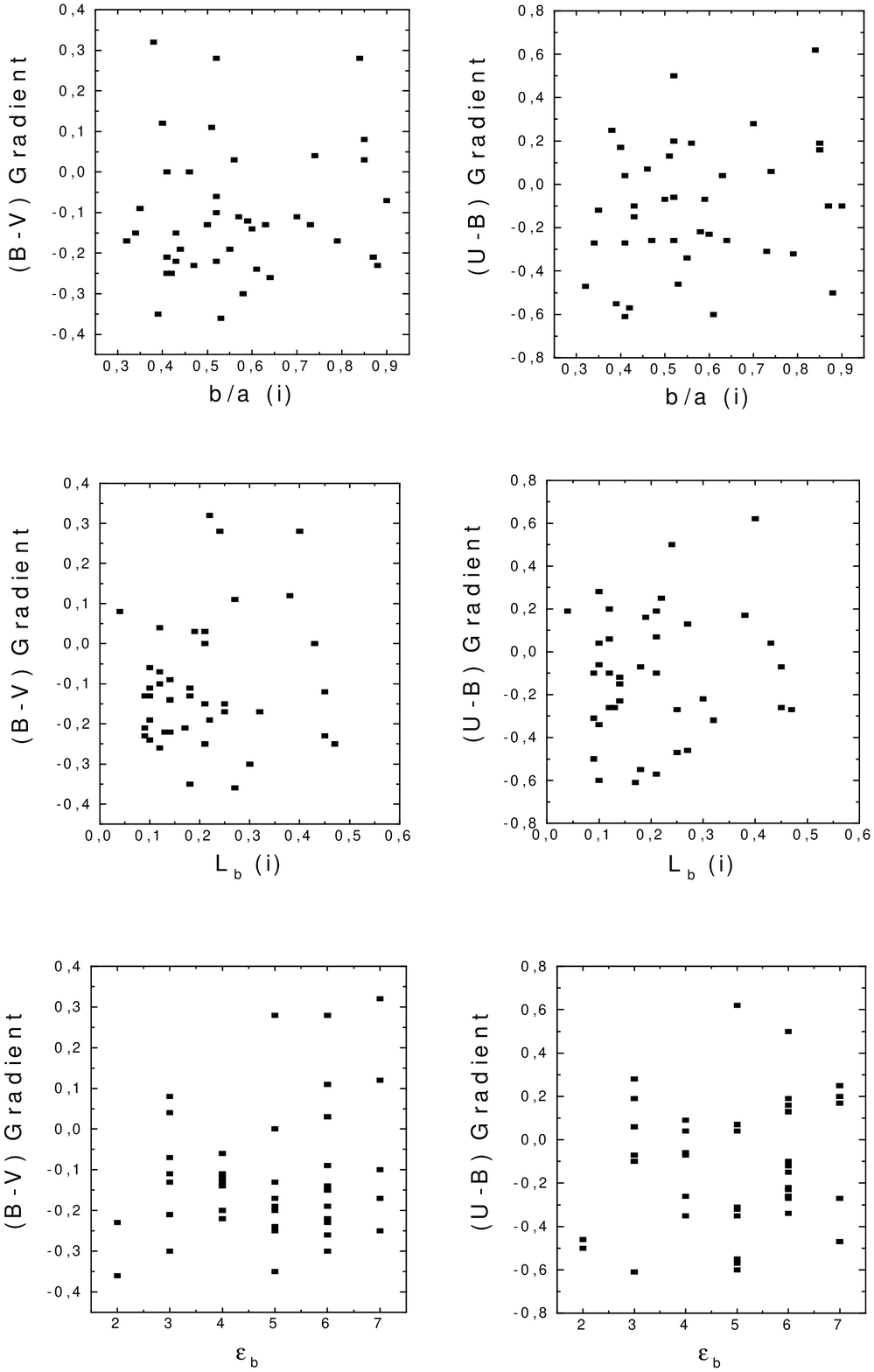}

\plotone{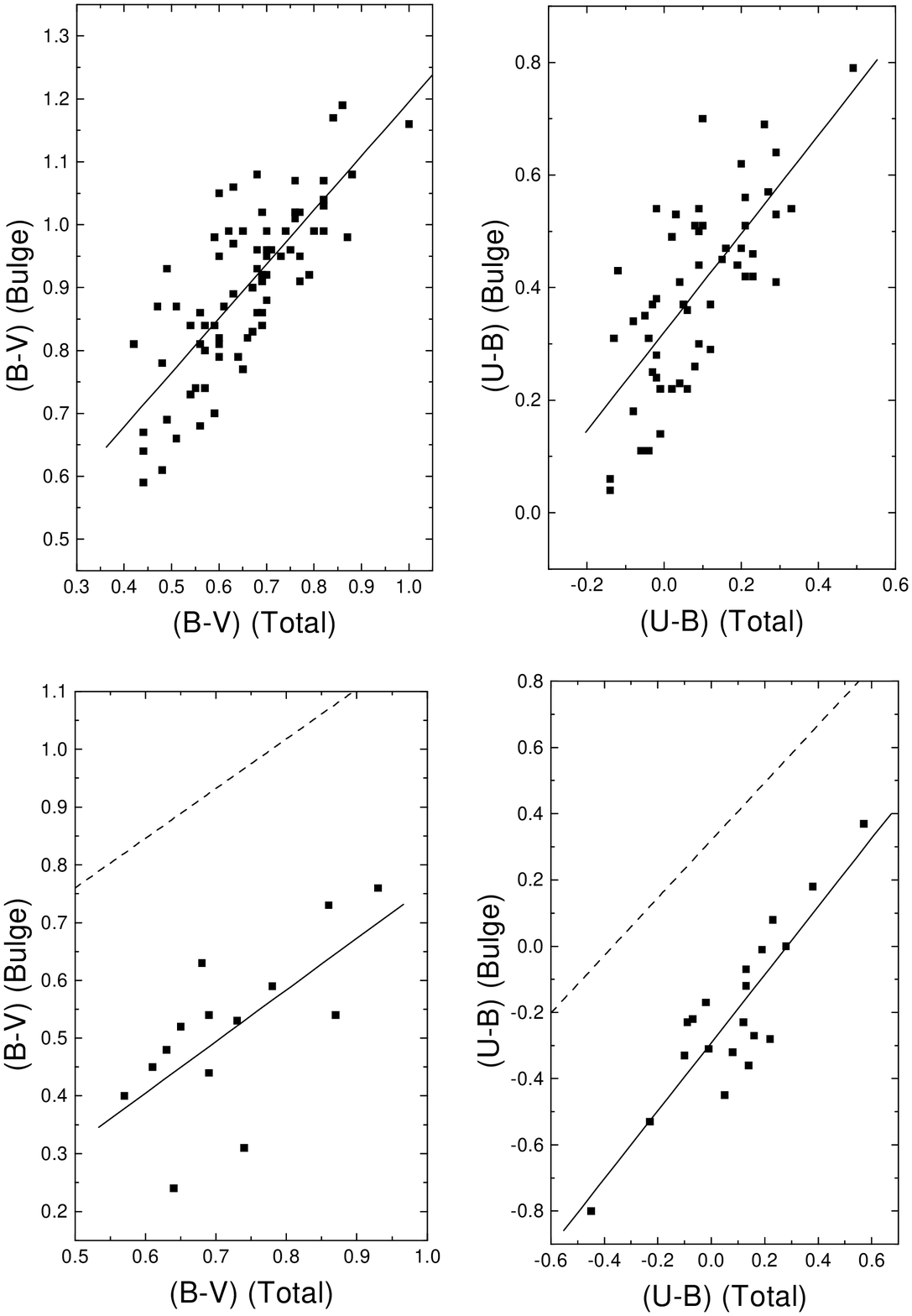}
 
\begin{deluxetable}{cccccccccc}
\tiny
\tablecaption{The sample. \label{tbl-1}}
\tablewidth{0pt}
\tablehead{
\colhead{Name} & \colhead{Type} & \colhead{$G(B-V)$} & \colhead{error} & \colhead{$G(U-B)$} & \colhead{error} & \colhead{$(B-V)_{b}$} & \colhead{$(B-V)_{T}$} & \colhead{$(U-B)_{b}$} & \colhead{$(U-B)_{T}$} \\ 
\colhead{(1)} & \colhead{(2)} & \colhead{(3)} & \colhead{(4)} & \colhead{(5)} & \colhead{(6)} & \colhead{(7)} & \colhead{(8)} & \colhead{(9)} & \colhead{(10)} \\ }
\startdata
ESO271-010	&	SABcd(s)	&	-0,02	&	0,03	&	 0,04	&	0,06	&	0,39	&	0,39	&	-0,22	&	-0,22
\\IC0342		&	SABcd(rs)	&	 0,43	&	0,08	&	 0,57	&	0,09	&	0,23	&	0,66	&	-0,51	&	 0,06
\\IC1954		&	SBb(s)		&	-0,06	&	0,03	&	 0,15	&	0,10	&	0,24	&	0,24	&	-0,40	&	-0,25
\\IC1993		&	SABb(rs)	&	-0,05	&	0,03	&	-0,01	&	0,10	&	0,73	&	0,73	&	 0,24	&	 0,24
\\IC2554		&	SBbc(s)		&	-0,06	&	0,03	&	 0,08	&	0,04	&	0,19	&	0,19	&	-0,45	&	-0,45
\\IC4444		&	SABbc(rs)	&	 0,00	&	0,03	&	 --	&	--	&	0,40	&	0,40	&	 --	&	 --
\\IC4839		&	SAbc(s)		&	-0,28	&	0,03	&	-0,44	&	0,06	&	0,88	&	0,60	&	 0,41	&	-0,03
\\IC4845		&	Sb(rs)		&	 0,13	&	0,00	&	-0,01	&	0,07	&	0,45	&	0,58	&	 0,05	&	 0,05
\\IC4852		&	SBbc(s)		&	-0,13	&	0,05	&	 0,03	&	0,07	&	0,67	&	0,54	&	-0,07	&	-0,07
\\IC5092		&	SBc(rs)		&	-0,01	&	0,03	&	-0,02	&	0,09	&	0,69	&	0,69	&	 0,09	&	 0,09
\\IC5179		&	Sbc(rs)		&	-0,17	&	0,04	&	-0,11	&	0,03	&	0,46	&	0,29	&	-0,12	&	-0,23
\\IC5186		&	SABb(rs)	&	-0,11	&	0,01	&	-0,11	&	0,04	&	0,53	&	0,42	&	-0,06	&	-0,17
\\IC5325		&	SABbc(rs)	&	-0,18	&	0,03	&	-0,05	&	0,05	&	0,69	&	0,51	&	-0,07	&	-0,07
\\MCG-2-14-4	&	SABcd(rs)	&	-0,33	&	0,11	&	 0,07	&	0,01	&	0,64	&	0,31	&	-0,11	&	-0,11
\\NGC0001		&	Sb		&	-0,14	&	0,01	&	-0,19	&	0,05	&	0,76	&	0,62	&	 0,19	&	 0,00
\\NGC0024		&	Sc(s)		&	-0,20	&	0,06	&	-0,14	&	0,06	&	0,29	&	0,09	&	-0,33	&	-0,47
\\NGC0134		&	SABbc(s)	&	-0,18	&	0,01	&	-0,35	&	0,03	&	0,49	&	0,31	&	 0,12	&	-0,23
\\NGC0150		&	SBb(rs)		&	-0,22	&	0,00	&	-0,21	&	0,04	&	0,54	&	0,32	&	-0,05	&	-0,26
\\NGC0151		&	SBbc(r)		&	-0,36	&	0,03	&	-0,46	&	0,06	&	0,73	&	0,37	&	 0,33	&	-0,13
\\NGC0157		&	SABbc(rs)	&	-0,23	&	0,03	&	-0,27	&	0,04	&	0,60	&	0,37	&	 0,07	&	-0,20
\\NGC0210		&	SABb(s)		&	-0,17	&	0,02	&	-0,32	&	0,00	&	0,78	&	0,61	&	 0,30	&	-0,02
\\NGC0224		&	Sb(s)		&	-0,03	&	0,00	&	-0,08	&	0,01	&      -0,10	&      -0,10	&	-0,26	&	-0,26
\\NGC0278		&	SABb(rs)	&	-0,03	&	0,04	&	 --	&	--	&	0,51	&	0,51	&	 --	&	 --
\\NGC0289		&	SBbc(rs)	&	-0,19	&	0,02	&	-0,34	&	0,02	&	0,78	&	0,59	&	 0,32	&	-0,02
\\NGC0309		&	SABc(r)		&	-0,63	&	0,04	&	-0,47	&	0,10	&	0,91	&	0,28	&	 0,26	&	-0,21
\\NGC0440		&	Sbc(s)		&	-0,16	&	0,02	&	-0,24	&	0,06	&	0,51	&	0,35	&	-0,03	&	-0,27
\\NGC0470		&	Sb(rs)		&	-0,13	&	0,03	&	 0,01	&	0,06	&	0,63	&	0,50	&	-0,06	&	-0,06
\\NGC0488		&	Sb(r)		&	-0,13	&	0,02	&	-0,28	&	0,04	&	0,90	&	0,77	&	 0,58	&	 0,30 \\
NGC0578		&	SABc(rs)	&	-0,21	&	0,03	&	-0,10	&	0,01	&	0,50	&	0,29	&	-0,15	&	-0,25
\\NGC0613		&	SBbc(rs)	&	 0,01	&	0,02	&	 0,03	&	0,04	&	0,63	&	0,63	&	 0,06	&	 0,06
\\ 
NGC0615		&	Sb(rs)		&	-0,12	&	0,01	&	-0,27	&	0,02	&	0,55	&	0,43	&	 0,29	&	 0,02
\\NGC0628		&	Sc(s)		&	-0,14	&	0,01	&	-0,23	&	0,02	&	0,64	&	0,50	&	 0,09	&	-0,14
\\NGC0685		&	SABc(r)		&	-0,22	&	0,03	&	-0,15	&	0,04	&	0,62	&	0,40	&	-0,04	&	-0,19
\\NGC0779		&	SABb(r)		&	-0,11	&	0,02	&	-0,25	&	0,03	&	0,45	&	0,34	&	 0,10	&	-0,16
\\NGC0782		&	SBb(r)		&	-0,31	&	0,00	&	-0,53	&	0,06	&	0,82	&	0,51	&	 0,44	&	-0,09
\\NGC0864		&	SABc(rs)	&	-0,11	&	0,03	&	 --	&	--	&	0,55	&	0,44	&	 --	&	 --
\\NGC0908		&	Sc(s)		&	-0,22	&	0,02	&	-0,50	&	0,04	&	0,54	&	0,32	&	 0,17	&	-0,33
\\NGC0958		&	SBc(rs)		&	-0,13	&	0,04	&	 --	&	--	&	0,48	&	0,35	&	 --	&	 --
\\NGC1055		&	SBb		&	-0,10	&	0,02	&	-0,26	&	0,04	&	0,53	&	0,43	&	 0,13	&	-0,13
\\NGC1068		&	Sb(rs)		&	-0,10	&	0,01	&	 0,01	&	0,01	&	0,74	&	0,64	&	 0,00	&	 0,00
\\NGC1073		&	SBc(rs)		&	-0,17	&	0,03	&	-0,47	&	0,04	&	0,62	&	0,45	&	 0,22	&	-0,25
\\NGC1084		&	Sc(s)		&	-0,14	&	0,04	&	-0,26	&	0,01	&	0,50	&	0,36	&	 0,03	&	-0,23
\\NGC1087		&	SABc(rs)	&	-0,01	&	0,04	&	 0,11	&	0,04	&	0,31	&	0,31	&	-0,36	&	-0,25
\\NGC1097		&	SBb(s)		&	 0,08	&	0,03	&	 0,12	&	0,02	&	0,64	&	0,64	&	 0,03	&	 0,15
\\NGC1187		&	SBc(r)		&	-0,15	&	0,02	&	-0,10	&	0,04	&	0,60	&	0,45	&	-0,01	&	-0,11
\\NGC1232		&	SABc(rs)	&	-0,24	&	0,01	&	-0,60	&	0,04	&	0,79	&	0,55	&	 0,45	&	-0,15
\\NGC1255		&	SABbc(rs)	&	-0,36	&	0,02	&	-0,14	&	0,04	&	0,62	&	0,26	&	-0,10	&	-0,24
\\NGC1288		&	SABc(rs)	&	-0,30	&	0,02	&	 --	&	--	&	0,88	&	0,58	&	 --	&	 --
\\NGC1300		&	SBbc(rs)	&	-0,18	&	0,03	&	-0,22	&	0,03	&	0,71	&	0,53	&	 0,21	&	-0,01
\\NGC1365		&	SBb(s)		&	 0,11	&	0,02	&	 0,13	&	0,04	&	0,43	&	0,54	&	-0,16	&	-0,03
\\NGC1421		&	SABbc(rs)	&	-0,30	&	0,02	&	-0,22	&	0,03	&	0,24	&      -0,06	&	-0,28	&	-0,50
\\NGC1425		&	Sb(s)		&	-0,13	&	0,02	&	-0,25	&	0,02	&	0,51	&	0,38	&	 0,13	&	-0,12
\\NGC1483		&	SBbc(s)		&	-0,09	&	0,04	&	-0,04	&	0,04	&	0,39	&	0,39	&	-0,26	&	-0,26
\\NGC1515		&	SABbc(s)	&	-0,09	&	0,02	&	-0,12	&	0,01	&	0,33	&	0,33	&	-0,05	&	-0,17
\\NGC1530		&	SBb(rs)		&	-0,12	&	0,04	&	-0,02	&	0,01	&	0,55	&	0,43	&	-0,11	&	-0,11
\\NGC1536		&	SBc(s)		&	-0,03	&	0,10	&	 --	&	--	&	0,51	&	0,51	&	 --	&	 --
\\NGC1566		&	SABbc(s)	&	-0,06	&	0,03	&	-0,06	&	0,05	&	0,57	&	0,57	&	-0,04	&	-0,04
\\NGC1614		&	SBc(s)		&	 0,02	&	0,03	&	 0,16	&	0,04	&	0,44	&	0,44	&	-0,34	&	-0,18
\\NGC1620		&	SABbc(rs)	&	-0,24	&	0,01	&	 --	&	--	&	0,56	&	0,32	&	 --	&	 --
\\NGC1637		&	SABc(rs)	&	-0,22	&	0,04	&	-0,26	&	0,06	&	0,76	&	0,54	&	 0,18	&	-0,08
\\NGC1672		&	SBb(s)		&	 0,00	&	0,01	&	 0,04	&	0,01	&	0,57	&	0,57	&	-0,05	&	-0,05
\\\tablebreak
NGC1688		&	SBd(rs)		&	 0,15	&	0,04	&	 0,42	&	0,12	&	0,30	&	0,45	&	-0,44	&	-0,02
\\NGC1703		&	SBb(r)		&	-0,18	&	0,00	&	-0,64	&	0,00	&	0,69	&	0,51	&	 0,36	&	-0,28
\\NGC1784		&	SBc(r)		&	-0,30	&	0,00	&	-0,22	&	0,05	&	0,81	&	0,51	&	 0,24	&	 0,02
\\NGC1792		&	Sbc(rs)		&	-0,19	&	0,03	&	-0,22	&	0,04	&	0,51	&	0,32	&	-0,06	&	-0,28
\\NGC1796		&	SBc(rs)		&	-0,08	&	0,01	&	-0,09	&	0,04	&	0,31	&	0,31	&	-0,26	&	-0,26
\\NGC1832		&	SBbc(r)		&	-0,43	&	0,02	&	-0,42	&	0,06	&	0,77	&	0,34	&	 0,19	&	-0,23
\\NGC1888		&	SBc(s)		&	-0,14	&	0,05	&	-0,26	&	0,05	&	0,43	&	0,29	&	 0,11	&	-0,15
\\NGC1961		&	SABc(rs)	&	-0,33	&	0,09	&	 --	&	--	&	0,73	&	0,40	&	 --	&	 --
\\NGC2082		&	SBb(r)		&	-0,18	&	0,03	&	-0,13	&	0,06	&	0,67	&	0,49	&	-0,08	&	-0,21
\\NGC2090		&	Sc(rs)		&	-0,10	&	0,02	&	-0,20	&	0,03	&	0,61	&	0,51	&	 0,17	&	-0,03
\\NGC2206		&	SABbc(rs)	&	-0,25	&	0,05	&	 --	&	--	&	0,66	&	0,41	&	 --	&	 --
\\NGC2207		&	SABbc(rs)	&	-0,32	&	0,03	&	 --	&	--	&	0,64	&	0,32	&	 --	&	 --
\\NGC2223		&	SABb(r)		&	-0,25	&	0,02	&	-0,57	&	0,08	&	0,88	&	0,63	&	 0,54	&	-0,03
\\NGC2268		&	SABbc(r)	&	-0,08	&	0,02	&	 --	&	--	&	0,53	&	0,53	&	 --	&	 --
\\NGC2336		&	SABbc(r)	&	-0,21	&	0,03	&	-0,61	&	0,05	&	0,60	&	0,39	&	 0,37	&	-0,24
\\NGC2339		&	SABbc(rs)	&	-0,12	&	0,02	&	 --	&	--	&	0,75	&	0,63	&	 --	&	 --
\\NGC2347		&	Sb(r)		&	-0,19	&	0,03	&	 --	&	--	&	0,76	&	0,57	&	 --	&	 --
\\NGC2389		&	SABc(rs)	&	-0,21	&	0,08	&	 --	&	--	&	0,55	&	0,34	&	 --	&	 --
\\NGC2417		&	Sbc(rs)		&	-0,12	&	0,03	&	-0,09	&	0,08	&	0,62	&	0,50	&	 0,00	&	 0,00 \\
NGC2442		&	SABbc(s)	&	-0,30	&	0,02	&	-0,30	&	0,01	&	0,86	&	0,56	&	 0,34	&	 0,04
\\NGC2487		&	SBb		&	-0,38	&	0,03	&	-0,42	&	0,08	&	0,93	&	0,55	&	 0,41	&	-0,01
\\NGC2512		&	SBb		&	-0,08	&	0,01	&	 0,02	&	0,05	&	0,54	&	0,54	&	-0,05	&	-0,05
\\NGC2565		&	SBbc		&	 0,02	&	0,01	&	-0,08	&	0,12	&	0,52	&	0,52	&	 0,07	&	 0,07
\\NGC2595		&	SABc(rs)	&	-0,06	&	0,02	&	-0,01	&	0,02	&	0,60	&	0,60	&	 0,07	&	 0,07
\\NGC2608		&	SBb(s)		&	-0,09	&	0,06	&	-0,08	&	0,07	&	0,53	&	0,53	&	-0,06	&	-0,06
\\NGC2613		&	Sb(s)		&	-0,16	&	0,01	&	-0,42	&	0,04	&	0,47	&	0,31	&	 0,28	&	-0,14
\\NGC2683		&	Sb(rs)		&	-0,08	&	0,01	&	-0,23	&	0,02	&	0,34	&	0,34	&	 0,07	&	-0,16
\\NGC2712		&	SBb(r)		&	-0,19	&	0,03	&	 --	&	--	&	0,63	&	0,44	&	 --	&	 --
\\NGC2715		&	SABc(rs)	&	-0,11	&	0,06	&	 --	&	--	&	0,27	&	0,16	&	 --	&	 --
\\NGC2776		&	SABc(rs)	&	 0,04	&	0,05	&	 0,04	&	0,05	&	0,52	&	0,52	&	-0,11	&	-0,11
\\NGC2815		&	SBb(r)		&	-0,24	&	0,02	&	-0,34	&	0,07	&	0,60	&	0,36	&	 0,37	&	 0,03
\\ 
NGC2841		&	Sb(r)		&	-0,12	&	0,01	&	-0,24	&	0,03	&	0,68	&	0,56	&	 0,36	&	 0,12
\\NGC2874		&	SBbc(r)		&	-0,26	&	0,04	&	-0,40	&	0,04	&	0,62	&	0,36	&	 0,23	&	-0,17
\\NGC2889		&	SABc(rs)	&	-0,26	&	0,02	&	-0,39	&	0,10	&	0,87	&	0,61	&	 0,37	&	-0,02
\\NGC2903		&	SABbc(rs)	&	 0,03	&	0,02	&	 0,16	&	0,03	&	0,40	&	0,40	&	-0,24	&	-0,08
\\NGC2935		&	SABb(s)		&	-0,12	&	0,05	&	-0,07	&	0,10	&	0,76	&	0,64	&	 0,13	&	 0,13
\\NGC2955		&	Sb(r)		&	 0,19	&	0,07	&	 --	&	--	&	0,21	&	0,40	&	 --	&	 --
\\NGC2964		&	SABbc(r)	&	-0,08	&	0,04	&	-0,06	&	0,01	&	0,49	&	0,49	&	-0,18	&	-0,18
\\NGC2989		&	SABbc(s)	&	-0,17	&	0,03	&	-0,12	&	0,00	&	0,42	&	0,25	&	-0,16	&	-0,28
\\NGC2997		&	SABc(rs)	&	 0,08	&	0,02	&	 0,19	&	0,03	&	0,68	&	0,68	&	 0,01	&	 0,20
\\NGC3001		&	SABbc(rs)	&	-0,09	&	0,06	&	-0,12	&	0,06	&	0,58	&	0,58	&	 0,02	&	-0,10
\\NGC3054		&	SABb(r)		&	-0,14	&	0,02	&	-0,23	&	0,03	&	0,74	&	0,60	&	 0,26	&	 0,03
\\NGC3079		&	SBc(s)		&	-0,25	&	0,01	&	-0,50	&	0,07	&	0,31	&	0,06	&	 0,02	&	-0,48
\\NGC3095		&	SABc(rs)	&	-0,39	&	0,03	&	-0,44	&	0,05	&	0,67	&	0,38	&	 0,19	&	-0,25
\\NGC3124		&	SABbc(rs)	&	-0,19	&	0,03	&	 --	&	--	&	0,80	&	0,61	&	 --	&	 --
\\NGC3145		&	SBbc(rs)	&	-0,25	&	0,01	&	-0,12	&	0,12	&	0,72	&	0,47	&	 0,18	&	 0,06
\\NGC3177		&	Sb(rs)		&	 0,04	&	0,02	&	-0,21	&	0,05	&	0,54	&	0,54	&	 0,51	&	 0,30
\\NGC3223		&	Sb(s)		&	-0,25	&	0,00	&	-0,34	&	0,04	&	0,76	&	0,50	&	 0,36	&	 0,02
\\NGC3281		&	Sab(s)		&	-0,10	&	0,01	&	-0,19	&	0,06	&	0,74	&	0,64	&	 0,31	&	 0,12
\\NGC3289		&	SB0+(rs)	&	-0,04	&	0,03	&	-0,03	&	0,02	&	0,31	&	0,31	&	-0,01	&	-0,01
\\NGC3310		&	SABbc(r)	&	 0,00	&	0,01	&	 0,05	&	0,04	&	0,19	&	0,19	&	-0,54	&	-0,54
\\NGC3318		&	SABb(rs)	&	-0,34	&	0,02	&	-0,50	&	0,04	&	0,58	&	0,24	&	 0,13	&	-0,37
\\NGC3333		&	SABbc		&	-0,19	&	0,04	&	 0,13	&	0,15	&	0,20	&	0,01	&	-0,55	&	-0,42
\\NGC3347		&	SBb(rs)		&	 0,03	&	0,01	&	-0,07	&	0,02	&	0,61	&	0,61	&	 0,16	&	 0,16
\\NGC3351		&	SBb(r)		&	 0,03	&	0,02	&	 0,19	&	0,05	&	0,66	&	0,66	&	-0,03	&	 0,16
\\NGC3353		&	Sb		&	 0,13	&	0,06	&	 0,13	&	0,05	&	0,17	&	0,30	&	-0,57	&	-0,44
\\NGC3390		&	Sb		&	-0,13	&	0,02	&	-0,16	&	0,08	&	0,32	&	0,19	&	-0,06	&	-0,22
\\NGC3521		&	SABbc(rs)	&	-0,07	&	0,03	&	 --	&	--	&	0,52	&	0,52	&	 --	&	 --
\\NGC3627		&	SABb(s)		&	-0,20	&	0,02	&	-0,21	&	0,02	&	0,62	&	0,42	&	 0,18	&	-0,03
\\NGC3628		&	Sb		&	-0,11	&	0,04	&	 --	&	--	&	0,28	&	0,17	&	 --	&	 --
\\NGC3689		&	SABc(rs)	&	 0,13	&	0,10	&	 --	&	--	&	0,37	&	0,50	&	 --	&	 --
\\NGC3810		&	Sc(rs)		&	-0,21	&	0,06	&	-0,27	&	0,07	&	0,64	&	0,43	&	 0,08	&	-0,19
\\\tablebreak
NGC4051		&	SABbc(rs)	&	 0,28	&	0,04	&	 0,50	&	0,05	&	0,37	&	0,65	&	-0,44	&	 0,06
\\NGC4088		&	SABbc(rs)	&	-0,36	&	0,01	&	 --	&	--	&	0,55	&	0,19	&	 --	&	 --
\\NGC4096		&	SABc(rs)	&	-0,20	&	0,05	&	-0,22	&	0,03	&	0,32	&	0,12	&	-0,15	&	-0,37
\\NGC4156		&	SBb(rs)		&	-0,01	&	0,08	&	 0,32	&	0,17	&	0,72	&	0,72	&	-0,31	&	 0,01
\\NGC4216		&	SABb(s)		&	-0,20	&	0,02	&	-0,29	&	0,03	&	0,59	&	0,39	&	 0,35	&	 0,06
\\NGC4254		&	Sc(s)		&	-0,15	&	0,01	&	-0,14	&	0,00	&	0,64	&	0,49	&	 0,06	&	-0,08
\\NGC4258		&	SABbc(s)	&	-0,06	&	0,03	&	 0,03	&	0,06	&	0,39	&	0,39	&	-0,04	&	-0,04
\\NGC4273		&	SBc(s)		&	-0,03	&	0,09	&	 --	&	--	&	0,33	&	0,33	&	 --	&	 --
\\NGC4303		&	SABbc(rs)	&	-0,13	&	0,03	&	 0,04	&	0,07	&	0,65	&	0,52	&	-0,01	&	-0,01
\\NGC4321		&	SABbc(s)	&	 0,04	&	0,01	&	 0,06	&	0,06	&	0,61	&	0,61	&	-0,04	&	-0,04
\\NGC4388		&	Sb(s)		&	-0,03	&	0,01	&	 0,09	&	0,01	&	0,17	&	0,17	&	-0,34	&	-0,34
\\NGC4414		&	Sc(rs)		&	-0,17	&	0,01	&	-0,37	&	0,15	&	0,75	&	0,58	&	 0,38	&	 0,01
\\NGC4501		&	Sb(rs)		&	-0,24	&	0,01	&	-0,34	&	0,03	&	0,76	&	0,52	&	 0,36	&	 0,02
\\NGC4527		&	SABbc(s)	&	-0,24	&	0,03	&	-0,14	&	0,05	&	0,68	&	0,44	&	 0,08	&	-0,06
\\NGC4535		&	SABc(s)		&	 0,09	&	0,02	&	 0,07	&	0,09	&	0,53	&	0,53	&	-0,06	&	-0,06
\\NGC4536		&	SABbc(rs)	&	-0,29	&	0,02	&	-0,17	&	0,06	&	0,54	&	0,25	&	-0,08	&	-0,25
\\NGC4548		&	SBb(rs)		&	-0,16	&	0,03	&	-0,28	&	0,01	&	0,87	&	0,71	&	 0,49	&	 0,21
\\NGC4565		&	Sb		&	-0,19	&	0,02	&	-0,20	&	0,07	&	0,32	&	0,13	&	 0,06	&	-0,14
\\NGC4579		&	SABb(rs)	&	-0,12	&	0,01	&	-0,16	&	0,02	&	0,84	&	0,72	&	 0,41	&	 0,25
\\NGC4593		&	SBb(rs)		&	 0,28	&	0,11	&	 0,62	&	0,03	&	0,65	&	0,93	&	-0,40	&	 0,22
\\NGC4647		&	SABc(rs)	&	-0,32	&	0,12	&	 --	&	--	&	0,83	&	0,51	&	 --	&	 --
\\NGC4651		&	Sc(rs)		&	-0,26	&	0,00	&	-0,52	&	0,00	&	0,73	&	0,47	&	 0,34	&	-0,18
\\NGC4666		&	SABc		&	-0,15	&	0,00	&	 --	&	--	&	0,42	&	0,27	&	 --	&	 --
\\NGC4699		&	SABb(rs)	&	-0,05	&	0,01	&	-0,14	&	0,02	&	0,75	&	0,75	&	 0,41	&	 0,27
\\NGC4900		&	SBc(rs)		&	 0,12	&	0,01	&	 0,12	&	0,09	&	0,42	&	0,54	&	-0,27	&	-0,15
\\NGC4902		&	SBb(r)		&	-0,39	&	0,09	&	 --	&	--	&	0,94	&	0,56	&	 --	&	 --
\\NGC4911		&	SABbc(r)	&	-0,17	&	0,06	&	-0,34	&	0,01	&	0,86	&	0,69	&	 0,41	&	 0,07
\\NGC4939		&	Sbc(s)		&	-0,29	&	0,02	&	-0,36	&	0,06	&	0,70	&	0,41	&	 0,24	&	-0,12
\\NGC5005		&	SABbc(rs)	&	-0,16	&	0,03	&	-0,18	&	0,02	&	0,69	&	0,53	&	 0,29	&	 0,11
\\NGC5033		&	Sc(s)		&	-0,30	&	0,03	&	-0,11	&	0,06	&	0,65	&	0,35	&	 0,15	&	 0,04
\\NGC5055		&	Sbc(rs)		&	-0,16	&	0,01	&	-0,37	&	0,06	&	0,68	&	0,52	&	 0,19	&	-0,18
\\ 
NGC5188		&	SABb(s)		&	-0,21	&	0,10	&	-0,34	&	0,06	&	0,62	&	0,41	&	 0,19	&	-0,15
\\NGC5194		&	Sbc(s)		&	-0,14	&	0,01	&	-0,24	&	0,02	&	0,58	&	0,44	&	 0,04	&	-0,20
\\NGC5236		&	SABc(s)		&	 0,32	&	0,01	&	 0,25	&	0,04	&	0,47	&	0,79	&	-0,20	&	 0,05
\\NGC5248		&	SABbc(rs)	&	-0,07	&	0,02	&	-0,10	&	0,03	&	0,58	&	0,58	&	 0,06	&	-0,04
\\NGC5364		&	Sbc(rs)		&	-0,18	&	0,02	&	-0,36	&	0,03	&	0,67	&	0,49	&	 0,24	&	-0,12
\\NGC5371		&	SABbc(rs)	&	-0,35	&	0,03	&	-0,55	&	0,04	&	0,91	&	0,56	&	 0,51	&	-0,04
\\NGC5426		&	Sc(s)		&	-0,22	&	0,04	&	-0,33	&	0,10	&	0,53	&	0,31	&	 0,00	&	-0,33
\\NGC5427		&	Sc(s)		&	-0,09	&	0,00	&	 --	&	--	&	0,52	&	0,52	&	 --	&	 --
\\NGC5483		&	Sc(s)		&	-0,14	&	0,03	&	 --	&	--	&	0,63	&	0,49	&	 --	&	 --
\\NGC5530		&	Sbc(rs)		&	-0,29	&	0,04	&	-0,50	&	0,02	&	0,74	&	0,45	&	 0,50	&	 0,00
\\NGC5592		&	SBbc(s)		&	-0,08	&	0,01	&	-0,05	&	0,05	&	0,52	&	0,52	&	-0,04	&	-0,04
\\NGC5633		&	Sb(rs)		&	-0,32	&	0,15	&	 --	&	--	&	0,70	&	0,38	&	 --	&	 --
\\NGC5643		&	SABc(rs)	&	 0,07	&	0,14	&	 0,21	&	0,08	&	0,58	&	0,58	&	-0,10	&	 0,11
\\NGC5653		&	Sb(rs)		&	 0,31	&	0,10	&	-0,09	&	0,07	&	0,24	&	0,55	&	-0,12	&	-0,12
\\NGC5676		&	Sbc(rs)		&	-0,22	&	0,02	&	-0,40	&	0,02	&	0,59	&	0,37	&	 0,20	&	-0,20
\\NGC5746		&	SABb(rs)	&	-0,17	&	0,04	&	-0,45	&	0,10	&	0,46	&	0,29	&	 0,29	&	-0,16
\\NGC5792		&	SBb(rs)		&	-0,21	&	0,02	&	-0,20	&	0,00	&	0,47	&	0,26	&	 0,01	&	-0,19
\\NGC5850		&	SBb(r)		&	-0,23	&	0,04	&	-0,26	&	0,05	&	0,90	&	0,67	&	 0,45	&	 0,19
\\NGC5859		&	SBbc(s)		&	-0,05	&	0,06	&	 --	&	--	&	0,40	&	0,40	&	 --	&	 --
\\NGC5861		&	SABc(rs)	&	-0,26	&	0,06	&	 --	&	--	&	0,68	&	0,42	&	 --	&	 --
\\NGC5879		&	Sbc(rs)		&	-0,14	&	0,02	&	 --	&	--	&	0,36	&	0,22	&	 --	&	 --
\\NGC5899		&	SABc(rs)	&	-0,18	&	0,06	&	 --	&	--	&	0,59	&	0,41	&	 --	&	 --
\\NGC5907		&	Sc(s)		&	-0,23	&	0,01	&	-0,30	&	0,04	&	0,17	&      -0,06	&	-0,21	&	-0,51
\\NGC5921		&	SBbc(r)		&	-0,15	&	0,02	&	-0,27	&	0,02	&	0,73	&	0,58	&	 0,23	&	-0,04
\\NGC5962		&	Sc(r)		&	-0,19	&	0,04	&	-0,16	&	0,06	&	0,66	&	0,47	&	 0,11	&	-0,05
\\NGC5970		&	SBc(r)		&	-0,17	&	0,00	&	 --	&	--	&	0,71	&	0,54	&	 --	&	 --
\\NGC5985		&	SABb(r)		&	-0,13	&	0,02	&	-0,31	&	0,03	&	0,67	&	0,54	&	 0,27	&	-0,04
\\NGC5987		&	Sb		&	-0,10	&	0,04	&	 --	&	--	&	0,63	&	0,53	&	 --	&	 --
\\NGC6052		&	Sc		&	 0,08	&	0,03	&	-0,16	&	0,03	&	0,18	&	0,18	&	 0,20	&	-0,54
\\NGC6181		&	SABc(rs)	&	-0,18	&	0,02	&	-0,17	&	0,03	&	0,49	&	0,31	&	-0,12	&	-0,29
\\NGC6207		&	Sc(s)		&	 0,06	&	0,02	&	-0,04	&	0,02	&	0,20	&	0,20	&	-0,39	&	-0,39
\\\tablebreak
NGC6217		&	SBbc(rs)	&	 0,12	&	0,04	&	 0,17	&	0,03	&	0,40	&	0,52	&	-0,31	&	-0,14
\\NGC6221		&	SBc(s)		&	-0,18	&	0,02	&	 0,03	&	0,03	&	0,62	&	0,44	&	-0,06	&	-0,06
\\NGC6239		&	SBb(s)		&	-0,05	&	0,02	&	 0,06	&	0,03	&	0,18	&	0,18	&	-0,39	&	-0,39
\\NGC6384		&	SABbc(r)	&	-0,26	&	0,03	&	-0,26	&	0,04	&	0,70	&	0,44	&	 0,28	&	 0,02
\\NGC6412		&	Sc(s)		&	-0,12	&	0,03	&	 --	&	--	&	0,57	&	0,45	&	 --	&	 --
\\NGC6574		&	SABbc(rs)	&	-0,12	&	0,03	&	-0,12	&	0,06	&	0,67	&	0,55	&	 0,10	&	-0,02
\\NGC6643		&	Sc(rs)		&	-0,14	&	0,01	&	-0,17	&	0,05	&	0,49	&	0,35	&	-0,10	&	-0,27
\\NGC6699		&	SABbc(rs)	&	-0,22	&	0,02	&	-0,24	&	0,04	&	0,76	&	0,54	&	 0,14	&	-0,10
\\NGC6744		&	SABbc(r)	&	-0,10	&	0,03	&	 0,20	&	0,09	&	0,71	&	0,61	&	 0,40	&	 0,60
\\NGC6753		&	Sb(r)		&	-0,13	&	0,02	&	-0,17	&	0,03	&	0,86	&	0,73	&	 0,24	&	 0,07
\\NGC6764		&	SBbc(s)		&	 0,13	&	0,06	&	 0,26	&	0,06	&	0,32	&	0,45	&	-0,35	&	-0,09
\\NGC6769		&	SABb(r)		&	-0,20	&	0,03	&	-0,49	&	0,13	&	0,76	&	0,56	&	 0,42	&	-0,07
\\NGC6780		&	SABc(rs)	&	-0,28	&	0,10	&	-0,04	&	0,07	&	0,76	&	0,48	&	 0,01	&	 0,01
\\NGC6814		&	SABbc(rs)	&	-0,11	&	0,05	&	 0,28	&	0,13	&	0,82	&	0,71	&	-0,06	&	 0,22
\\NGC6872		&	SBb(s)		&	-0,11	&	0,02	&	-0,24	&	0,03	&	0,52	&	0,39	&	 0,31	&	 0,07
\\NGC6887		&	Sbc		&	-0,20	&	0,07	&	-0,37	&	0,09	&	0,52	&	0,32	&	 0,16	&	-0,21
\\NGC6890		&	Sb(rs)		&	-0,21	&	0,02	&	-0,24	&	0,04	&	0,82	&	0,61	&	 0,24	&	 0,00
\\NGC6923		&	SBb(rs)		&	-0,38	&	0,04	&	-0,33	&	0,04	&	0,76	&	0,38	&	 0,23	&	-0,10
\\NGC6925		&	Sbc(s)		&	-0,35	&	0,02	&	-0,52	&	0,12	&	0,55	&	0,20	&	 0,23	&	-0,29
\\NGC6951		&	SABbc(rs)	&	 0,00	&	0,05	&	 0,07	&	0,07	&	0,62	&	0,62	&	 0,10	&	 0,10
\\NGC6984		&	SBc(r)		&	-0,35	&	0,03	&	-0,36	&	0,01	&	0,62	&	0,27	&	 0,06	&	-0,30
\\NGC7038		&	SABc(s)		&	-0,13	&	0,03	&	-0,27	&	0,03	&	0,63	&	0,50	&	 0,25	&	-0,02
\\NGC7083		&	Sbc(s)		&	-0,20	&	0,02	&	-0,27	&	0,02	&	0,63	&	0,43	&	 0,13	&	-0,14
\\NGC7090		&	SBc		&	-0,01	&	0,06	&	-0,05	&	0,01	&      -0,11	&      -0,11	&	-0,59	&	-0,59
\\NGC7125		&	SABc(rs)	&	-0,14	&	0,01	&	 0,09	&	0,09	&	0,36	&	0,22	&	-0,20	&	-0,20
\\NGC7126		&	Sc(rs)		&	-0,17	&	0,02	&	-0,30	&	0,05	&	0,49	&	0,32	&	 0,01	&	-0,29
\\NGC7137		&	SABc(rs)	&	 0,02	&	0,01	&	 0,07	&	0,02	&	0,51	&	0,51	&	-0,12	&	-0,12
\\NGC7171		&	SBb(rs)		&	-0,20	&	0,01	&	-0,35	&	0,05	&	0,67	&	0,47	&	 0,15	&	-0,20
\\NGC7177		&	SABb(r)		&	-0,16	&	0,02	&	-0,17	&	0,02	&	0,76	&	0,60	&	 0,33	&	 0,16
\\NGC7184		&	SBc(r)		&	-0,28	&	0,01	&	-0,33	&	0,09	&	0,50	&	0,22	&	 0,11	&	-0,22
\\NGC7205		&	Sbc(s)		&	-0,29	&	0,01	&	-0,44	&	0,02	&	0,60	&	0,31	&	 0,11	&	-0,33
\\ 
NGC7314		&	SABbc(rs)	&	-0,21	&	0,02	&	-0,32	&	0,05	&	0,54	&	0,33	&	 0,01	&	-0,31
\\NGC7329		&	SBb(r)		&	-0,20	&	0,04	&	-0,35	&	0,01	&	0,78	&	0,58	&	 0,38	&	 0,03
\\NGC7331		&	Sb(s)		&	-0,13	&	0,01	&	-0,33	&	0,04	&	0,57	&	0,44	&	 0,25	&	-0,08
\\NGC7339		&	SABbc(s)	&	 0,08	&	0,11	&	 0,17	&	0,03	&	0,32	&	0,32	&	-0,37	&	-0,20
\\NGC7412		&	SBb(s)		&	-0,23	&	0,01	&	-0,50	&	0,03	&	0,64	&	0,41	&	 0,29	&	-0,21
\\NGC7448		&	Sbc(rs)		&	-0,14	&	0,02	&	-0,22	&	0,00	&	0,30	&	0,16	&	-0,17	&	-0,39
\\NGC7479		&	SBc(s)		&	-0,25	&	0,02	&	-0,27	&	0,02	&	0,78	&	0,53	&	 0,30	&	 0,03
\\NGC7496		&	SBb(s)		&	 0,23	&	0,03	&	 0,33	&	0,06	&	0,35	&	0,58	&	-0,42	&	-0,09
\\NGC7531		&	SABbc(r)	&	-0,23	&	0,01	&	-0,38	&	0,02	&	0,60	&	0,37	&	 0,19	&	-0,19
\\NGC7537		&	Sbc		&	-0,22	&	0,03	&	-0,08	&	0,04	&	0,30	&	0,08	&	-0,42	&	-0,42
\\NGC7541		&	SBbc(rs)	&	-0,21	&	0,04	&	-0,22	&	0,05	&	0,45	&	0,24	&	-0,10	&	-0,32
\\NGC7590		&	Sbc(rs)		&	-0,36	&	0,03	&	-0,37	&	0,10	&	0,55	&	0,19	&	 0,05	&	-0,32
\\NGC7606		&	Sb(s)		&	-0,18	&	0,04	&	-0,29	&	0,08	&	0,57	&	0,39	&	 0,11	&	-0,18
\\NGC7640		&	SBc(s)		&	-0,20	&	0,02	&	-0,01	&	0,02	&	0,02	&      -0,18	&	-0,59	&	-0,59
\\NGC7673		&	Sc		&	 0,12	&	0,04	&	 0,23	&	0,14	&	0,19	&	0,31	&	-0,58	&	-0,35
\\NGC7716		&	SABb(r)		&	-0,22	&	0,03	&	 --	&	--	&	0,81	&	0,59	&	 --	&	 --
\\NGC7723		&	SBb(r)		&	-0,09	&	0,04	&	 0,00	&	0,05	&	0,58	&	0,58	&	 0,00	&	 0,00
\\NGC7742		&	Sb(r)		&	-0,15	&	0,05	&	-0,37	&	0,20	&	0,79	&	0,63	&	 0,32	&	-0,05
\\NGC7755		&	SBc(rs)		&	-0,13	&	0,03	&	-0,07	&	0,04	&	0,74	&	0,61	&	 0,05	&	 0,05
\\NGC7757		&	Sc(rs)		&	-0,24	&	0,07	&	-0,29	&	0,12	&	0,51	&	0,27	&	-0,07	&	-0,36
\\NGC7782		&	Sb(s)		&	-0,25	&	0,06	&	-0,59	&	0,19	&	0,82	&	0,56	&	 0,51	&	-0,08
\\UGC03973	&	SBb		&	 0,13	&	0,10	&	 0,30	&	0,07	&	0,30	&	0,43	&	-1,02	&	-0,72
\\UGC04013	&	Sb		&	 0,13	&	0,05	&	 0,44	&	0,12	&	0,13	&	0,26	&	-1,08	&	-0,64
\\\tablecomments{Columns (1) and (2) show, respectively, the name and the morphological type of the
galaxy, according to the RC3. Columns (3) and (4) show the (B\,$-$V) gradient and its error, while
columns (5) and (6) do that for the (U$-$B) gradient. Columns (7) and (8) show the (B\,$-$V) bulge
and total color indices, respectively, whereas columns (9) and (10) do that for the (U$-$B) gradient.}
\enddata
\end{deluxetable}

\clearpage
 
\begin{deluxetable}{ccccccc}
\normalsize
\tablecaption{Total number of galaxies, mean values and standard deviations for the distributions
presented in Fig. 5. \label{tbl-2}}
\tablewidth{0pt}
\tablehead{
\colhead{Description} & \colhead{N} & \colhead{mean value} & \colhead{SD} & \colhead{N} & \colhead{mean value} & \colhead{SD} \\ 
\colhead{(1)} & \colhead{(2)} & \colhead{(3)} & \colhead{(4)} & \colhead{(5)} & \colhead{(6)} & \colhead{(7)} \\ 
\colhead{} & \colhead{} & \colhead{G(B\,$-$V)} & \colhead{} & \colhead{} & \colhead{G(U$-$B)} & \colhead{}}
\startdata
(a) non--barred face--on galaxies		& 26   	& -0.14 $\pm$ 0.01    & 0.06   & 22      & -0.19 $\pm$ 0.03  & 0.14\nl
(b) barred face--on galaxies			& 98  	& -0.14 $\pm$ 0.02    & 0.15   & 82      & -0.08 $\pm$ 0.03  & 0.27\nl
(c) non--barred edge--on galaxies		& 46   	& -0.16 $\pm$ 0.01    & 0.07   & 41      & -0.26 $\pm$ 0.02  & 0.12\nl
(d) barred edge--on galaxies			& 68   	& -0.16 $\pm$ 0.01    & 0.10   & 56      & -0.19 $\pm$ 0.03  & 0.20\nl
\enddata
\end{deluxetable}

\clearpage

\begin{deluxetable}{ccccccccc}
\small
\tablecaption{Distribution of the face--on galaxies in our sample in relation to the gradient categories to be
analysed. \label{tbl-3}}
\tablewidth{0pt}
\tablehead{
\colhead{} & \colhead{Color} & \colhead{Total} & \colhead{Sample} & \colhead{SA} & \colhead{SAB} & \colhead{SB} & \colhead{SAB+SB} & \colhead{AGN} \\ 
\colhead{} & \colhead{} & \colhead{(1)} & \colhead{(2)} & \colhead{(3)} & \colhead{(4)} & \colhead{(5)} & \colhead{(6)} & \colhead{(7)}}
\startdata
G $\geq$ 0.1              & (B\,-V) & 14    & 11\%    & 28\% & 29\% & 43\% & \hskip0.4cm72\%     & 5(36\%)\\
-0.1 $<$ G $<$ 0.1        & 	    & 32    & 26\%    &  9\% & 53\% & 38\% & \hskip0.4cm91\%     & 4(12\%)\\ 
G $\leq$ -0.1             & 	    & 78    & 63\%    & 25\% & 47\% & 28\% & \hskip0.4cm75\%     & 8(10\%)\\
G $\geq$ 0.1         	  & (U-B)   & 19    & 18\%    & 10\% & 37\% & 53\% & \hskip0.4cm90\%     & 7(37\%)\\ 
-0.1 $<$ G $<$ 0.1        & 	    & 30    & 29\%    & 17\% & 53\% & 30\% & \hskip0.4cm83\%     & 2(7\%)\\ 
G $\leq$ -0.1             & 	    & 55    & 53\%    & 27\% & 42\% & 31\% & \hskip0.4cm73\%     & 4(7\%)\\
\tablecomments{(1): total number of galaxies in the category; (2): fraction of the total sample 
in the category; (3): fraction of non--barred galaxies; (4): fraction of weakly--barred galaxies; 
(5): fraction of barred galaxies; (6): total fraction of barred galaxies; (7): galaxies with AGN.}
\enddata
\end{deluxetable}

\clearpage
 
\begin{deluxetable}{ccccccc}
\normalsize
\tablecaption{Median values for the total and bulge characteristic color indices for the galaxies
in our sample, separated in relation to the gradient categories. The values for the whole sample
are shown in the left part of the table, while those for the face--on objects are in the right. 
\label{tbl-4}}
\tablewidth{0pt}
\tablehead{
\colhead{} & \colhead{Color} & \colhead{Bulge} & \colhead{Total} & \colhead{Color} & \colhead{Bulge} & \colhead{Total} \\ 
\colhead{} & \colhead{} & \colhead{Total Sample} & \colhead{} & \colhead{} & \colhead{Face--on} & \colhead{}}
\startdata
G $\geq$ 0.1          & (B\,$-$V)	& 0.34$\pm$0.03  	  & 0.53$\pm$0.04 	 & (B\,$-$V)      &  0.36$\pm$0.03  & 0.55$\pm$0.05 \\
-0.1 $<$ G $<$ 0.1    &	 	   	& 0.52$\pm$0.03           & 0.52$\pm$0.03        &  	          &  0.57$\pm$0.02  & 0.57$\pm$0.02 \\ 
G $\leq$ -0.1         &	 		& 0.64$\pm$0.01  	  & 0.43$\pm$0.01        &  	          &  0.74$\pm$0.01  & 0.53$\pm$0.01 \\
G $\geq$ 0.1          & (U$-$B) 	& -0.35$\pm$0.06 	  & -0.08$\pm$0.05	 &  (U$-$B)       &  -0.31$\pm$0.07 & 0.05$\pm$0.07  \\ 
-0.1 $<$ G $<$ 0.1    &	 		& -0.06$\pm$0.03          & -0.06$\pm$0.03       &  	          &  -0.05$\pm$0.03 & -0.05$\pm$0.03 \\ 
G $\leq$ -0.1         &  		& 0.19$\pm$0.02  	  & -0.14$\pm$0.01	 &  	          &  0.24$\pm$0.03  & -0.05$\pm$0.02 \\
\enddata
\end{deluxetable}

\clearpage

\begin{deluxetable}{cccccccccccccccc}
\footnotesize
\tablecaption{Color gradients and excesses for the galaxies in the dust extinction studies.
\label{tbl-5}}
\tablewidth{0pt}
\tablehead{
\colhead{Galaxy} & \colhead{G(B-V)} & \colhead{G(B-I)} & \colhead{G(B-H)} & \colhead{G(B-K)} & \colhead{G(V-I)} & \colhead{G(V-H)} &
\colhead{G(V-K)} \\ \colhead{G(I-H)} & \colhead{G(I-K)} & \colhead{G(H-K)} & \colhead{E(V-I)} & \colhead{E(V-H)} & \colhead{E(V-K)} & \colhead{E(I-H)} & \colhead{E(I-K)}}
\startdata
\bf{NGC 3310} & & -0.41 & & & & & \\ -0.11 & -0.11 & -0.04 & & & & & 0.42 \\
\bf{NGC 5033} & +0.01 & & +0.39 & & & +0.01 & \\ & & & & & & & \\
\bf{NGC 5194} & -0.40 & & -0.50 & -0.47 & & & -0.21 \\ & & -0.02 & & & 1.41 & & \\
\bf{NGC 5248} & & & & & +0.20 & -0.24 & \\ -0.38 & & & & 0.68 & & 0.75 & \\
\bf{NGC 782}  & -0.36 & & & & -0.10 & & \\ & & & 0.11 & & & & \\
\bf{NGC 6769} & +0.16 & & & & -0.18 & & \\ & & & 0.25 & & & & \\
\bf{NGC 6890} & -0.19 & & & & -0.06 & & \\ & & & 0.04 & & & & \\
\bf{NGC 6923} & -0.37 & & & & -0.28 & & \\ & & & 0.31 & & & & \\
\bf{NGC 7496} & +0.31 & & & & -0.30 & & \\ & & & 0.22 & & & & \\
\enddata
\end{deluxetable}

\clearpage

\begin{deluxetable}{ccccc}
\footnotesize
\tablecaption{Comparison between the color gradients with (right) and without (left) internal
extinction correction.
\label{tbl-6}}
\tablewidth{0pt}
\tablehead{
\colhead{Galaxy} & \colhead{$G(B-V)_0$} & \colhead{$G(B-V)_c$} & \colhead{$G(U-B)_0$} & \colhead{$G(U-B)_c$}}
\startdata
NGC 3310 & 0.00 & +0.26 & +0.05 & 0.35 \\
NGC 5194 & -0.14 & -0.12 & -0.24 & +0.09 \\
NGC 5248 & -0.07 & +0.11 & -0.10 & -0.05 \\
NGC 782  & -0.31 & -0.28 & -0.53 & -0.47 \\
NGC 6769 & -0.20 & -0.19 & -0.49 & -0.42 \\
NGC 6890 & -0.21 & -0.18 & -0.24 & -0.09 \\
NGC 6923 & -0.38 & -0.38 & -0.33 & -0.22 \\
NGC 7496 & +0.23 & +0.23 & +0.33 & +0.40 \\
\enddata
\end{deluxetable}


\clearpage

\begin{thebibliography}{}

   \bibitem[Athanassoula \& Bureau 1999]{ath99}
      Athanassoula, E., and Bureau, M. 1999, accepted for publication in \apj, astro-ph/9904206

   \bibitem[Baugh, Cole \& Frenk 1996]{bau96}
      Baugh, C.M., Cole, S., and Frenk, C.S. 1996, \mnras, 283, 1361

   \bibitem[Berentzen et al. 1998]{ber98}
      Berentzen, I., Heller, C.H., Shlosman, I., and Fricke, K.J. 1998, \mnras, 300, 49 

   \bibitem[Boselli \& Gavazzi 1994]{bos94}
      Boselli, A., and Gavazzi, G. 1994, \aap, 283, 12

   \bibitem[Bouwens, Cay\'on, \& Silk 1998]{bou98}
      Bouwens, R., Cay\'on, L., and Silk, J. 1998, astro-ph/9812193, accepted for publication 
      in \apj
 
   \bibitem[Bureau \& Athanassoula 1999]{bur99a}
      Bureau, M., and Athanassoula, E. 1999, accepted for publication in \apj, astro-ph/9903061

   \bibitem[Bureau \& Freeman 1999]{bur99b}
      Bureau, M., and Freeman, K.C. 1999, accepted for publication in \aj, astro-ph/9904015

   \bibitem[Bureau, Freeman \& Athanassoula 1999]{bur99c}
      Bureau, M., Freeman, K.C., and Athanassoula, E. 1999, in When and How do Bulges Form and
      Evolve?, ed. by C.M. Carollo, H.C. Ferguson \& R.F.G. Wyse, Cambridge: CUP, astro-ph/9901246

   \bibitem[Combes \& Sanders 1981]{com81}
      Combes, F., and Sanders, R.H. 1981, \aap, 96, 164 
   
   \bibitem[Courteau, de Jong \& Broeils (1996)]{cou96} 
      Courteau, S., de Jong, R., and Broeils, A. 1996, \apjl, 457, L73
   
   \bibitem[de Jong (1996a)]{dej96a}
      de Jong, R.S. 1996a, \aap, 118, 557
      
   \bibitem[de Jong (1996b)]{dej96b}
      de Jong, R.S. 1996b, \aap, 313, 45
   
   \bibitem[de Jong (1996c)]{dej96c}
      de Jong, R.S. 1996c, \aap, 313, 377

   \bibitem[de Jong \& van der Kruit (1994)]{dej94}
      de Jong, R.S., and van der Kruit, P.C. 1994, \aap, 106, 451

   \bibitem[de Souza \& dos Anjos 1987]{des87}
      de Souza, R.E., and dos Anjos, S. 1987, \aap, 70, 465

   \bibitem[G. de Vaucouleurs 1959]{dev59}
      de Vaucouleurs, G. 1959, \aj, 64, 397

   \bibitem[G. de Vaucouleurs et al. 1991]{dev91}
      de Vaucouleurs, G., de Vaucouleurs, A., Corwin, H.G., Buta, R.J., Paturel, G., 
      and Fouque P. 1991, in: Third Reference Catalog of Bright Galaxies, Springer--Verlag,
      New York {\bf (RC3)}
  
   \bibitem[Eggen, Lynden--Bell \& Sandage 1962]{egg62} 
      Eggen, O.J., Lynden--Bell, D., and Sandage, A.R. 1962, \apj, 136, 748

   \bibitem[Elmegreen (1998)]{elm98}
      Elmegreen, D.M. 1998, in Galaxies and Galactic Structure, Prentice Hall

   \bibitem[Evans 1994]{eva94}
      Evans, R. 1994, \mnras, 266, 511

   \bibitem[Friedli 1999]{fri99}
      Friedli, D. 1999, in The Evolution of Galaxies on Cosmological Timescales, 
      ed. by J.E. Beckman \& T.J. Mahoney, ASP Conf. Ser., astro-ph/9903143

   \bibitem[Friedli \& Benz 1995]{fri95}
      Friedli, D., and Benz, W. 1995, \aap, 301, 649
   
   \bibitem[Friedli \& Martinet (1993)]{fri93}
      Friedli, D., and Martinet, L. 1993, \aap, 277, 27 

   \bibitem[Frogel 1985]{fro85}
      Frogel, J.A. 1985, \apj, 298, 528

   \bibitem[Giovanelli et al. (1994)]{gio94}
      Giovanelli, R., Haynes, M.P., Salzer, J.J., Wegner, G., da Costa, L.N., and Freudling, W., 
      1994, \aj, 107(6), 2036

   \bibitem[Giovanelli et al. 1995]{gio95}
      Giovanelli, R., Haynes, M.P., Salzer, J.J., Wegner, G., da Costa, L.N., and Freudling, W., 
      1995, \aj, 110(3), 1059

   \bibitem[Graham (1982)]{gra82}
      Graham, J.A. 1982, \pasp, 94, 244

   \bibitem[Jansen et al. 1994]{jan94}
      Jansen, R.A. et al. 1994, \mnras, 270, 343

   \bibitem[Kauffmann \& White 1993]{kau93}
      Kauffmann, G., and White, S.D.M. 1993, \mnras, 261, 921

   \bibitem[Kauffmann, Guiderdoni \& White 1994]{kau94}
      Kauffmann, G., Guiderdoni, B., and White, S.D.M. 1994, \mnras, 267, 981

   \bibitem[Kitchin (1998)]{kit98}
      Kitchin, C.R. 1998, in Astrophysical Techniques, Institute of Physics Publishing, 
      Bristol and Philadelphia

   \bibitem[Kormendy (1982)]{kor82}
      Kormendy, J. 1982, \apj, 257, 75

   \bibitem[Kormendy \& Illingworth (1983)]{kor83}
      Kormendy, J., and Illingworth, G. 1983, \apj, 265, 632

   \bibitem[Kuijken \& Merrifield 1995]{kui95}
      Kuijken, K., and Merrifield, M.R. 1995, \apj, 443, L13

   \bibitem[Lahav et al. 1995]{lah95}
      Lahav, O., Naim, A, Buta, R.J., Corwin, H.G., and de Vaucouleurs, G. et al. 1995, Science, 267,
      859
   
   \bibitem[Larson \& Tinsley (1978)]{lar78}
      Larson, R.B., and Tinsley, B.M. 1978, \apj, 219, 46

   \bibitem[Longo and A. de Vaucouleurs (1983)]{lon83} 
      Longo, G., and de Vaucouleurs, A. 1983, Univ. Texas Monographs in Astronomy, No. 3 {\bf (LdV83)}
      
   \bibitem[Longo \& A. de Vaucouleurs 1985]{lon85}
      Longo, G., and de Vaucouleurs, A. 1985, Univ. Texas Monographs in Astronomy, No. 3A {\bf (LdV85)}
      
   \bibitem[Martin (1995)]{mar95}
      Martin, P. 1995, \aj, 109(6), 2428 {\bf (M95)}
      
   \bibitem[Martin \& Roy (1994)]{mar94}
      Martin, P., and Roy, J.R. 1994, \apj, 424, 599 {\bf (MR94)}

   \bibitem[Merrifield \& Kuijken 1999]{mer99}
      Merrifield, M.R., and Kuijken, K. 1999, \aap, 345, L47

   \bibitem[Norman, Sellwood \& Hasan (1996)]{nor96}
      Norman, C.A., Sellwood, J.A., and Hasan, H. 1996, \apj, 462, 114
      
   \bibitem[Peletier 1989]{pel89}
      Peletier, R.F. 1989, PhD Thesis, University of Groningen, The Netherlands

   \bibitem[Peletier \& Balcells 1996]{pel96}
      Peletier, R.F., and Balcells, M. 1996, \aj, 111, 2238
      
   \bibitem[Peletier et al. 1994]{pel94}
      Peletier, R.F., Valentjin, E.A., Moorwood, A.F.M., and Freudling, W. 1994, \aap, 108, 621

   \bibitem[Peletier et al. 1995]{pel95}
      Peletier, R.F. et al. 1995, \aap, 300, L1

   \bibitem[Peletier et al. 1999]{pel99}
      Peletier, R.F. et al. 1999, \mnras, 310, 703

   \bibitem[Prugniel \& H\'eraudeau (1998)]{pru98}
      Prugniel, Ph., and H\'eraudeau, Ph. 1998, \aap, 128, 299 {\bf (PH98)}

   \bibitem[Rieke \& Lebofsky (1985)]{rie85}
      Rieke, G., Lebofsky, M.J. 1985, \apj, 288, 618

   \bibitem[Roberts \& Haynes 1994]{rob94}
      Roberts, M.S., and Haynes, M.P. 1994, \araa, 32, 115
   
   \bibitem[Rousseeuw 1984]{rou84}
      Rousseeuw, P.J. 1984, Journal of the American Statistical Association, 79(388), 871
      
   \bibitem[Rousseeuw \& Leroy 1987]{rou87}
      Rousseeuw, P.J., and Leroy, A.M. 1987, in: Robust Regression and Outlier Detection,
      Wiley--Interscience, New York
   
   \bibitem[Sakamoto et al. (1999)]{sak99}
      Sakamoto, K., Okumura, S.K., Ishizuki, S., and Scoville, N.Z. 1999, 
      in When and How do Bulges Form and Evolve?, ed. by C.M. Carollo, H.C. Ferguson \&
      R.F.G. Wyse, Cambridge University Press, astro-ph/9902005

   \bibitem[Schlegel, Finkbeiner \& Davis (1998)]{sch98}
      Schlegel, D.J., Finkbeiner, D.P., and Davis, M. 1998, \apj, 500, 525
   
   \bibitem[Shlosman, Begelman \& Frank 1990]{shl90}
      Shlosman, I., Begelman, M.C., and Frank, J. 1990, \nat, 345, 679

   \bibitem[Shlosman, Frank \& Begelman 1989]{shl89}
      Shlosman, I., Frank, J., and Begelman, M.C. 1989, \nat, 338, 45

   \bibitem[Searle, Sargent \& Bagnuolo 1973]{sea73}
      Searle, L., Sargent, W.L.W., and Bagnuolo, W.G. 1973, \apj, 179, 427

   \bibitem[Shaw 1987]{sha87}
      Shaw, M.A. 1987, \mnras, 229, 691

   \bibitem[Silva \& Elston 1994]{sil94}
      Silva, D.R., and Elston, R. 1994, \apj, 428, 511

   \bibitem[Tinsley 1980]{tin80}
      Tinsley, B.M. 1980, Fund. of Cos. Phys., 5, 287

   \bibitem[Toomre 1966]{too66}
      Toomre, A. 1966, in Geophysical Fluid Dynamics, 1966 Summer Study Program at Woods
      Hole Oceanographic Institution, ref. no. 66-46, 111
   
   \bibitem[van den Bergh 1997]{van97}
      van den Bergh, S. 1997, \aj, 113, 2054
  
   \bibitem[V\'eron--Cetty \& V\'eron (1998)]{ver98}
      V\'eron--Cetty, M.P., and V\'eron, P. 1998, in: Quasars and Active Galactic Nuclei (8th Ed.),
      ESO Sci. Rep., 18, 1
   
   \bibitem[Wyse, Gilmore \& Franx 1997]{wys97}
      Wyse, R.F.G., Gilmore, G., and Franx, M. 1997, \araa, 35, 637
  
   \bibitem[Zaritsky, Kennicutt \& Huchra (1994)]{zar94}
      Zaritsky, D., Kennicutt, R.C., and Huchra, J.P. 1994, \apj, 420, 87 {\bf (ZKH94)}
   
\end{thebibliography}
\end{document}